\documentclass[12pt]{article}

\usepackage[a4paper,margin=1in]{geometry}
\usepackage{setspace}
\usepackage{amsmath,amssymb,amsfonts}
\usepackage{amsthm}
\usepackage{bm}
\usepackage{mathrsfs}

\usepackage{graphicx}
\usepackage{booktabs}
\usepackage{multirow}
\usepackage{url}

\usepackage{algorithm}
\usepackage{algpseudocode}
\usepackage{listings}
\usepackage{xcolor}

\usepackage{natbib}

\usepackage[title]{appendix}

\theoremstyle{definition}

\title{Informed Asymmetric Dirichlet Priors for Multivariate Bernoulli Mixture Models}

\author{
Luisa Ferrari\thanks{Department of Economics, University of Modena and Reggio Emilia, Italy. Email: \texttt{luisaferrari@unimore.it}} \\
\and
Maria Franco Villoria\thanks{Department of Economics, University of Modena and Reggio Emilia, Italy.} \\
\and
Garritt L. Page\thanks{Department of Statistics, Brigham Young University, Provo, UT, USA.} \\
\and
Alex Laini\thanks{Department of Life Sciences, University of Turin, Italy.}
}

\date{}

\begin{document}

\maketitle

\begin{abstract}
Clustering multivariate binary data is of interest in many scientific fields, including ecology, biomedicine, and social policy. Beyond heuristic clustering algorithms, such data can be modelled using multivariate Bernoulli mixture models. Many Bayesian implementations of these models involve a trade-off between computational efficiency and full posterior inference.  We propose instead a Bayesian approach able to provide both aspects. The method fixes the total number of components to a large value and employs an asymmetric Dirichlet prior on the mixture weights. The asymmetric Dirichlet hyperparameters are elicited using the popular Penalized Complexity prior framework, which provides an intuitive way for users to inform the induced distribution of the number of clusters. An efficient MCMC algorithm is then developed to fit the model. Simulations and real-world applications demonstrate that the method is competitive with existing alternatives and can outperform them in certain settings. The proposal is illustrated using an ecological dataset about presence–absence of species across multiple sites, where cluster-specific parameters are modelled on the basis of environmental conditions. Overall, the proposed method provides a computationally efficient, fully Bayesian, and interpretable framework for clustering multivariate binary data, with potential applications across diverse scientific domains.\end{abstract}


\section{Introduction}

Multivariate binary data consist of multiple binary responses recorded jointly for the same statistical unit, so that each observation is a vector of 0/1 outcomes rather than a single binary variable. This structure arises naturally in many scientific fields. In biomedicine, individuals are frequently described by binary indicators representing the presence or absence of multiple diseases, clinical conditions or genetic variants, as in comorbidity studies, disease classification and treatment response \citep{whitson2016identifying}. In ecological studies, a species (or another taxon) is typically detected in only a subset of the sampled sites within a given area. In particular, studies that only record binary presence–absence information for species (or intraspecies genetic variants) have become increasingly common due to the widespread adoption of DNA metabarcoding techniques for genetic identification \citep{taberlet2018environmental}. In social policy, countries are often described by sets of binary indicators capturing the adoption of policies or institutional features \citep{tkacova2024welfare}.

In many instances, it is of interest to cluster the multivariate binary observations to find underlying structures. For example, clustering ecological presence–absence matrices can help identify groups of species with similar habitat preferences, revealing co-occurring species assemblages \citep{legendre2012numerical}. These assemblages are useful for informing conservation strategies and improving ecological monitoring by summarizing complex biodiversity patterns into fewer subgroups. Likewise, in biomedicine and genetics, clustering individuals based on binary disease or genetic profiles can uncover subphenotypes, which may correspond to distinct therapeutic responses or risk factors, thereby supporting the need for targeted interventions and personalized medicine \citep{abel1993autologistic,sun2007multivariate}. Similar objectives appear in several other fields, including text classification \citep{juan2002use}, voting data \citep{6298015}, and animal classification \citep{li2005general}. These applications illustrate the broad utility of clustering methods for multivariate binary data.

While algorithmic clustering procedures are often straightforward to implement, they are inferentially limited and challenging to extend to complex modeling scenarios. On the other hand, model-based approaches provide probabilistic assignment of observations to clusters, can readily incorporate covariates, easily accommodate missing data, quantify uncertainty in cluster parameter estimation, and facilitate estimation of the number of clusters from the data \citep{mclachlan2000finite}. Model-based clustering is typically formulated through finite mixture models (FMMs), in which the distribution of an observed response (denoted by $Y_i$) is expressed as the following convex combination of $K$ component distributions with mixing proportions $\omega_k\geq0,\sum_{k=1}^K \omega_k=1$:
\begin{align}\label{eq:basic_mixture}
p(Y_i~\mid~\bm{\theta}_1, \ldots, \bm{\theta}_K) = \sum_{k=1}^K \omega_k p_k(Y_i~\mid~\bm{\theta}_k).
\end{align}
Here, the $\bm{\theta}_k \in \mathbb{R}^d$ denote cluster-specific parameters.  The number of components $K$ can be fixed to a specific value or assigned a prior.  Either way however, $K$ does not necessarily correspond to the number of clusters, which is instead defined as the data-informed number of occupied components (denoted by $K^+$), and can be estimated regardless of how $K$ is treated. 

In the context of multivariate binary data, $p_k(\cdot ~\mid~ \bm{\theta}_k)$ in \eqref{eq:basic_mixture} is a multivariate Bernoulli distribution, where conditional independence is usually assumed between the binary variables within clusters \citep{carreira2000practical}.  There are a variety of software packages that implement multivariate Bernoulli mixture models, including the EM-based implementation available in \texttt{flexmix} \citep{leisch2004flexmix}, the MCMC-based Bayesian approach of \texttt{BayesBinMix} \citep{papastamoulis2017bayesbinmix}, and the more recent variational inference proposal of \texttt{VICatMix} \citep{rao2025vicatmix}. 

However, existing software implementations tend to prioritise either computational efficiency or full posterior inference, failing to achieve both simultaneously. Moreover, to the best of our knowledge, none of the available packages readily accommodate the inclusion of covariates in the modelling of latent component-specific parameters. An additional limitation, both in this context and more broadly in the FMM literature, is the little attention given to the specification of prior information on the number of clusters. As mentioned, in most cases the number of components $K$ is either fixed {\it a priori} or assigned a relatively uninformative prior, with little consideration about the induced prior on the number of occupied components $K^+$ and its implications in terms of prior beliefs and posterior inference.

To address this gap, we propose a novel prior specification for multivariate Bernoulli mixture models, building on the recent approach of \cite{page2025informed}, which has proven successful in univariate and multivariate Gaussian mixture models. The method is based on the use of an asymmetric Dirichlet prior on the mixture weights ($\omega_1, \ldots, \omega_K$), which enables efficient posterior inference while retaining a fully Bayesian formulation. This specification also provides the advantage of an intuitive and flexible way to incorporate prior information about the number of clusters. In particular, it allows the user to tune the prior through two hyperparameters: one representing a soft upper bound on the expected number of clusters, and another reflecting the degree of uncertainty regarding this upper bound relative to smaller values. Furthermore, this framework facilitates prior sensitivity analysis, as the prior is formulated in a transparent and consistent way. This contributes to the growing literature on user-friendly and interpretable prior specification \citep{hartmann2020flexible,fuglstad2020intuitive,mikkola2024prior}.

The remainder of the paper is organized as follows. In Section \ref{sec:mbmm}, we provide the necessary background and notation for multivariate Bernoulli mixture models, and review existing implementations from the literature. Section \ref{sec:proposal} presents our proposed method, based on an asymmetric Dirichlet prior for the mixture weights, along with the details of the MCMC implementation used for posterior inference. In Section \ref{sec:simulations}, we evaluate the performance of the proposed approach through a simulation study. Section \ref{sec:applications} illustrates the application of our method in two case studies: a simpler example on a handwritten digits dataset, and a real-world ecological application on an original dataset obtained from an author-led DNA metabarcoding campaign. Finally, Section \ref{sec:discussion} concludes with a discussion. Code to replicate the simulations and applications is available at the associated GitHub repository: \url{https://github.com/LFerrariIt/binary_clustering/}.

\section{Multivariate Bernoulli mixture model}\label{sec:mbmm}
In this section, we introduce the formal framework for multivariate Bernoulli mixture models (MBMMs), essential for the understanding of the rest of the paper.
Consider a set of $N$ observations. For each unit $i= 1, \ldots, N$, we consider a multivariate binary vector of dimension $P$ such that:
$$\mathbf{Y}_i = (Y_{i1}, \ldots, Y_{iP}), \qquad Y_{ip} \in \{0,1\},\quad p=1,\ldots,P.$$
If the $N$ observations are assumed to be generated by different distributions, it is convenient to model the data using a finite mixture of conditionally independent Bernoulli distributions with $K$ latent components or clusters. Thus, $\bm{\theta}_k$ in \eqref{eq:basic_mixture} correspond to probabilities which we denote by $\boldsymbol{\pi}_k=(\pi_{k1},\ldots,\pi_{kP})$. Marginally, the probability mass function of $\mathbf{Y}_i$ is given by:
\begin{equation}\label{eq:MBMM}
p(\mathbf{Y}_i~\mid~\boldsymbol{\omega},\boldsymbol{\pi}_1,\ldots,\boldsymbol{\pi}_K)
= \sum_{k=1}^K \omega_k\left[
\prod_{p=1}^P
\pi_{kp}^{Y_{ip}} (1 - \pi_{kp})^{1 - Y_{ip}}\right]
\end{equation}
where $\boldsymbol{\omega} = [\omega_1,\ldots,\omega_K]^T$ denotes the vector of mixture proportions and each $\pi_{kp} \in [0,1]$ represents the probability of success for dimension $p$ in mixture component $k$. 

In a Bayesian setting, the model requires specifying priors on the success probabilities. 
In the absence of clear prior information, Jeffreys priors are typically assumed which in our setting correspond to $\pi_{kp}\sim \text{Beta}(0.5,0.5)$ for $k=1,\ldots,K$ and $\;p=1,\ldots,P$.

In some applications, covariates are naturally associated with the $P$ dimensions of the response vector. This occurs, for instance, in ecological studies where units $i$ correspond to species and dimensions $p$ represent spatial locations: the sites may be characterized by environmental conditions (e.g., climate, topography, or habitat features) that can help explain systematic variation in species presence across locations. To accommodate this setting within the MBMM framework, we let $\mathbf{x}_p\in \mathbb{R}^q$ denote the covariate vector associated with dimension $p$: the success probabilities $\pi_{kp}$ can then be modelled through cluster-specific logistic regressions, i.e. $\text{logit}(\pi_{kp}) = \mathbf{x}_p^\top \boldsymbol{\beta}_k$. The regression parameters $\bm{\beta}_k$ can simply be assigned vague Gaussian priors.

An equivalent formulation of the model introduces a set of unknown latent allocation variables $Z_i \in \{1,\ldots,K\},$ for $ i=1,\ldots,N$, which denote the cluster membership of unit $i$. The latent variables $Z_i$ are assumed to follow a categorical distribution with probabilities given by the mixture weights $\boldsymbol{\omega}$:
$$Z_i \sim \text{Categorical}(\boldsymbol{\omega}),\quad i=1,\ldots,N.$$
Conditional on $Z_i = k$, the variables within $\mathbf{Y}_i$ are assumed to be conditionally independent Bernoulli random variables:
$$Y_{ip} \mid Z_i = k \sim \text{Bernoulli}(\pi_{kp}), \qquad  i=1,\ldots,N;\quad p = 1,\ldots,P.$$
For computational reasons, it is more convenient to think in terms of this second formulation and consider the joint likelihood of the observed $\mathbf{Y}_i$ and the unknown $Z_i$, assuming independence between the two.

In this model, the parameter $K$ denotes the number of components specified {\it a priori}. This quantity is often distinct from the number of clusters or non-empty components ($K^+$). If $K^+$ were known in advance, one could trivially set $K = K^+$. However, in most practical applications $K^+$ is unknown, and the choice of $K$ therefore requires careful consideration. Within the broader mixture modelling literature, several strategies have been proposed to address this issue. A commonly adopted approach is to fit the model for a range of candidate values of $K$ and to compare the resulting models using information criteria such as AIC, BIC, or ICL. This strategy is implemented, for example, in the \texttt{flexmix} package, which estimates general mixture models, including MBMMs, via an EM algorithm. While this approach is computationally efficient, it does not quantify posterior uncertainty about $K^+$ and consequently fails to propagate this uncertainty to the estimation of the remaining model parameters. In the Bayesian world, two separate strategies have been proposed for handling the unknown number of components $K$. The first is to assign a prior directly to $K$, yielding so-called mixtures of FMM models \citep{miller2018mixture}. The \texttt{BayesBinMix} package \citep{papastamoulis2017bayesbinmix} implements this strategy for MBMM estimation by assigning to $K$ either a discrete Uniform or a truncated Poisson prior, formally incorporating uncertainty about the number of clusters. The remaining parameters are specified using commonly adopted priors: component-specific probabilities are modelled as $\pi_{kp} \sim \text{Beta}(a,b)$, while the mixture weights are assigned a symmetric Dirichlet $\boldsymbol{\omega} \sim \text{Dirichlet}(\alpha,\ldots,\alpha)$. Estimation is carried out via an MCMC scheme with parallel tempering, in which multiple heated chains, corresponding to different temperature levels that flatten the posterior distribution,
exchange states to promote efficient mixing and exploration across models with different values of $K$, though at a substantial computational cost. 
A second approach found in the Bayesian literature, defined as the sparse FMM approach (sFMM) \cite{malsiner2016model}, fixes $K$ to a relatively large value (exceeding the expected value of $K^+$) and relies on a prior on the mixture weights $\boldsymbol{\omega}$ that shrinks some weights towards zero, typically by setting $\alpha<1$ in the symmetric Dirichlet prior. Other recent works also consider assigning a prior to $\alpha$ \citep{malsiner2016model, schnatter:2019, greve:2020, schnatter_etal:2021}. Recently, \cite{rao2025vicatmix} proposed a variational inference approach for fitting an sFMM to substantially reduce runtimes. The method finds a mean-field approximation to the true posterior that minimizes the Kullback–Leibler divergence \citep{kld-1951}, equivalently maximizing the Evidence Lower Bound (ELBO). Optimization is carried out via coordinate ascent: the variational distribution of the cluster labels is updated conditional on the current parameter distributions, followed by updating the parameter distributions conditional on the updated cluster-label distribution. Since the optimization is only guaranteed to converge to a local optimum, multiple random initializations are used. The resulting solutions are summarized into a final partition by constructing a co-clustering matrix across runs and selecting the partition that minimizes the variation of information (VI) criterion \citep{meilua2007comparing}. The method, implemented in the \texttt{VICatMix} R package, therefore provides only point estimates of the cluster labels.

Despite the range of methods, a gap remains in MBMM estimation: computationally efficient packages provide fast point estimates but limited uncertainty quantification, whereas the MCMC-based \texttt{BayesBinMix} delivers full posterior inference at a computational cost that is prohibitive for moderate to large datasets. Furthermore, none of the existing methods considers how to incorporate intuitively prior information on the expected number of clusters into the model.

\section{Prior for mixture weights}\label{sec:proposal}
Motivated by this gap in the literature, we investigate a combination of existing techniques for MBMMs that seeks to balance computational efficiency with posterior inference, while at the same time enabling intuitive and flexible prior elicitation. Specifically, we adopt the sparse FMM framework for faster estimation, incorporate a recently proposed prior specification to allow more control over the introduction of prior knowledge, and implement estimation via MCMC to retain full posterior uncertainty.

The starting point of the proposal requires fixing $K$ to a large value, i.e. larger than the expected number of clusters. The second step consists in choosing a prior on $\boldsymbol{\omega}$ such that the number of non-empty components $K^+$ is shrunk to smaller values than $K$. Instead of considering the typical symmetric Dirichlet, we employ an asymmetric Dirichlet prior developed for generic mixture models (aFMM) by \cite{page2025informed} to the MBMM context. Specifically, the technique assumes:
\begin{gather}\label{eq:aFMM}
    \boldsymbol{\omega} \sim \text{Dirichlet}(\underbrace{\alpha_1,\ldots,\alpha_1}_{U},
\underbrace{\alpha_2,\ldots,\alpha_2}_{K-U}),\quad \alpha_1>0,\alpha_2>0
\end{gather}
where $U$ represents the value to which it is desired to ``center" the induced prior on $K^+$.  $\alpha_1$ controls the filling of the first $U$ components, while $\alpha_2$ empties the last $K-U$ components. The induced prior on $K^+$ is more concentrated around $U$ as $\alpha_1\rightarrow\infty$ and $\alpha_2\rightarrow 0$. Fixing the \( \alpha_2 \) parameter to a small value ensures that $U$ becomes a soft upper bound for $K^+$, since the last \( K-U \) components are shrunk towards negligible weights and spurious clusters are discouraged. After some investigation, we found $\alpha_2 = 0.01$ to be a reasonable choice in the context of MBMMs.

With respect to \( \alpha_1 \), larger values correspond to stronger prior belief that the number of non-empty components \( K^+ \) is close to \( U \). In particular, setting \( \alpha_1 \) to values on the order of \( U \) leads to an increasingly concentrated prior for \( K^+ \) around \( U \). Instead of fixing \( \alpha_1 \) to a large value, we allow the data to inform its magnitude while still encouraging shrinkage toward \( U \). To this end, we introduce a prior on \( \alpha_1 \) that favours larger values up to \( U \), while retaining flexibility to deviate from this preference when supported by the data. As in \cite{page2025informed}, the Penalized Complexity (PC) framework \citep{simpson2017penalising} is used to assign a PC prior to \( \alpha_1 \) with base model \( U \), defined on the support \( (0, U) \):
$$\alpha_1 \sim PC_U(\lambda),\quad 0<\alpha_1<U$$
The hyperparameter \( \lambda \) is calibrated through a tail probability statement on the induced prior of \( K^+ \), namely $
P(K^+ < U) = tp$, where $tp \in (0,1)$ controls the amount of prior mass assigned below the target number of non-empty components. Thereby, the prior specification is governed in practice by two intuitive hyperparameters: \( U \) represents a soft upper bound for \( K^+ \); $tp$  controls the uncertainty by regulating the left-tail deviation from $U$. More details on the derivation of the PC prior in this setting can be found in \cite{page2025informed}.

\begin{figure}[H]
    \centering
    \includegraphics[width=\textwidth]{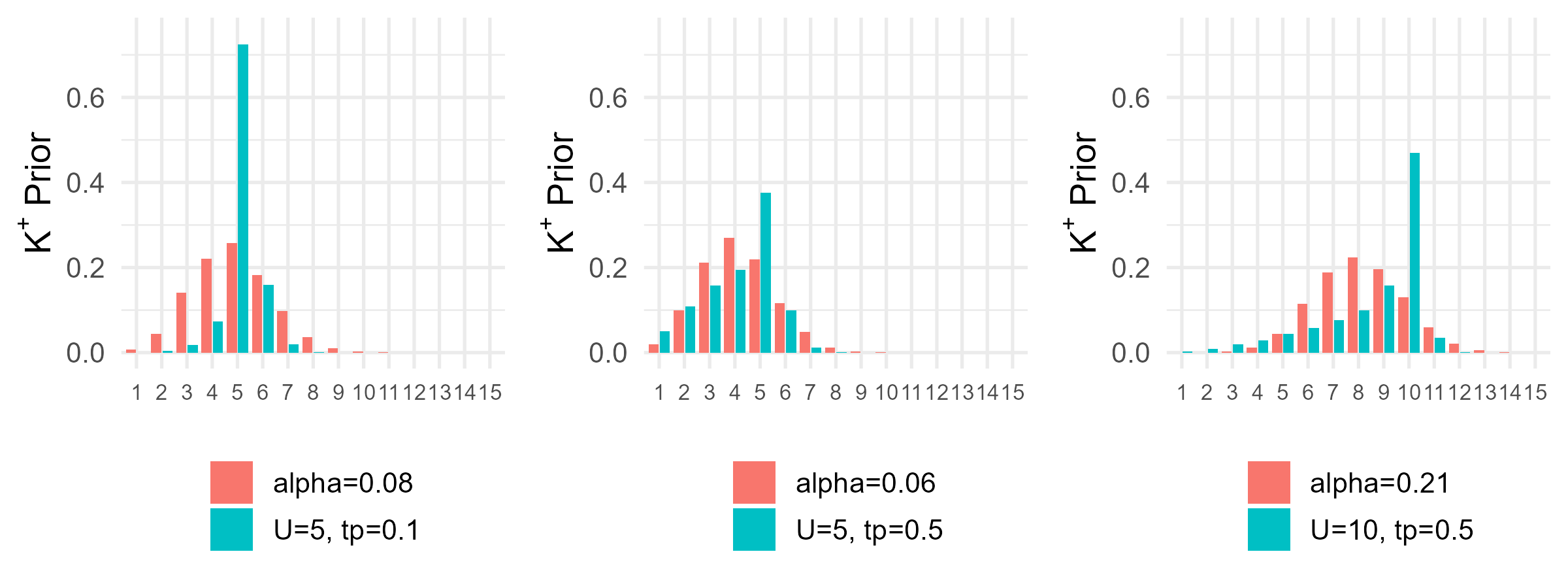}
    \caption{Induced prior distributions on $K^+$ when $N=100$ and $K=15$: the asymmetric Dirichlet prior (blue) with different choices of $U$ and $tp$; the symmetric Dirichlet prior (red) with the $\alpha$ value that minimizes the KLD from the asymmetric choice.} \label{fig:prior_plot}
\end{figure}

Figure \ref{fig:prior_plot} shows examples of the induced prior on $K^+$ under different aFMM prior specifications, compared with their symmetric Dirichlet counterparts, where the concentration parameter $\alpha$ (the common shape parameter of the symmetric Dirichlet prior) is chosen by numerically minimizing the Kullback–Leibler divergence between the induced prior distributions on $K^+$. The figure illustrates that the aFMM prior places a clear soft upper bound at $U$, with probability mass that progressively shifts toward smaller values of $K^+$ as $tp$ increases, thereby strengthening the left tail of the distribution. In contrast, the symmetric Dirichlet prior tends to concentrate its mass around a value that cannot be immediately controlled by the user through the parameter $\alpha$, and it does not allow for separate control of the distribution spread.

\subsection{MCMC implementation}

The proposed model is estimated using a single MCMC chain based on a block-Gibbs sampling scheme, in which \(\mathbf{z},\boldsymbol{\omega}, \boldsymbol{\pi}_{1:K}, \alpha_1\) are updated sequentially. All full conditional distributions are available in closed form, except for that of \(\alpha_1\), which is updated via a random-walk Metropolis--Hastings (MH) step. When covariates are included, additional MH updates are required for the regression parameters \(\boldsymbol{\beta}_k\). 

To accelerate convergence, the allocation vector \(\mathbf{z}\) is initialized using the \(k\)-modes algorithm \citep{huang1997fast}, as in \cite{rao2025vicatmix}. 
Despite this informed initialization, the allocation vector \(\mathbf{z}\) still often suffers from posterior multimodality, corresponding to alternative modal partitions. This multimodality can be problematic for MCMC sampling, as it may lead to poor mixing and slow exploration of the posterior space. In practice, addressing this issue is essential to ensure the algorithm is competitive with the non-MCMC approaches mentioned above.

To this end, we considered simulated annealing \citep{hajek1988cooling}, which consists in flattening a target function through exponentiation by a power less than one, yielding a so-called tempered (or heated) version that is easier to explore. In this context, the full conditional distribution of \(\mathbf{z}\) is flattened, reducing the separation between high- and low-probability allocations and facilitating transitions between alternative partitions. As a result, the chain is less prone to local trapping and explores the space of cluster assignments more effectively.
Specifically, at each iteration \(b\), \(\mathbf{z}\) is sampled from:
\[
p_T(z \mid \mathbf{Y}, \boldsymbol{\pi}_{1:K}, \boldsymbol{\omega})
\;\propto\;
p(z \mid \mathbf{Y}, \boldsymbol{\pi}_{1:K}, \boldsymbol{\omega})^{1/T_b},
\]
where \(T_b \ge 1\) denotes the temperature at iteration \(b\). In this context, we set \(T_1 = 5\) and decrease it to 1 over the first 90\% iterations. The sequence of temperatures follows an exponential decay, obtained by equally spacing the values on the logarithmic scale, and then mapping them back via exponentiation.
For the final 10\% iterations, the temperature is fixed at \(T_b = 1\), corresponding to standard sampling from the target conditional distribution. These iterations are retained for posterior inference after discarding the initial heated phase (burn-in).

Finally, the label-switching problem is addressed by enforcing an ordering constraint whenever inference on cluster-specific parameters is of interest: at each MCMC iteration, cluster labels and their corresponding parameters are relabelled so that clusters are indexed in decreasing order of size.

An illustrative implementation of the MCMC algorithm in {\tt R} is reported in the Supplementary Material.

\section{Simulation study}\label{sec:simulations} 

We conducted a simulation study to evaluate the performance of our proposed method (aFMM) against the most relevant competitors. 

The simulated data has been generated  as follows. We fixed $N=100$ and considered 6 different cases combining $P\in\{20,50\}$ and $K^+\in\{2,5,10\}$. Given the values of $P$ and $K^+$, we simulated the probabilities of success per cluster and per variable $\pi_{kp}$ with two different mechanisms: $\pi_{kp}\sim \text{Unif}(0,1)$ for Scenario 1; $\pi_{kp}\sim \text{Beta}\left(1/3,1\right)$ for Scenario 2, to replicate real-world cases where there is imbalance between the two classes. After fixing the success probabilities for each case, we created 50 datasets by first generating cluster labels $z_i\sim \text{DiscreteUnif}[1,K^+]$ and then responses $y_{ip} \sim \text{Bernoulli}(\pi_{z_i,p})$ for $i=1,\ldots,N;\;p=1,\ldots,P$.

The simulated data was then modelled using our proposal (aFMM), considering $U\in\{2,5,10\}$ to test sensitivity to prior choice, and setting $K=15$, $\alpha_2=0.01$, $tp=0.5$: the MCMC algorithm was run for 10,000 total iterations. As competitors, we considered \texttt{VICatMix} with $K=15$, and $\alpha\in\{0.01,0.1,0.5\}$ to induce different levels of shrinkage; we used the model averaging option of the method with 20 initializations. We also included a variant equivalent to \texttt{VICatMix} that uses a symmetric Dirichlet prior on the weights but estimated via MCMC, denoted here as sFMM, in order for us to be able to disentangle whether observed differences arise solely from the choice of estimation algorithm or also from the use of the asymmetric Dirichlet prior: the hyperparameters of sFMM are set as in \texttt{VICatMix}. For \texttt{flexmix}, we considered $K$ from 1 to 15, ten initializations for each model, and we then selected the best model using either AIC, BIC, or ICL. Finally, the \texttt{BayesBinMix} method was excluded due to its prohibitive computational cost, recently highlighted by the comparative study carried out in \cite{rao2025vicatmix}. 

To compare the methods, we considered two performance metrics. Firstly, we evaluated the bias in the estimation of the number of clusters. Secondly, we assessed clustering accuracy using the Adjusted Rand Index (ARI) between the true and estimated cluster labels.
As a point estimator for the partition for the MCMC-based methods, we used the \texttt{minVI} function from \texttt{VICatMix} for consistency with the main competitor, which finds the partition that minimizes the lower bound to the posterior expected Variation of Information \citep{meilua2007comparing}. Note that estimating the partition using instead the popular SALSO algorithm, introduced by \cite{dahl2022search} and implemented in the \texttt{salso R} package, yields very similar results.

\begin{figure}[H]
    \centering
    \includegraphics[width=\textwidth]{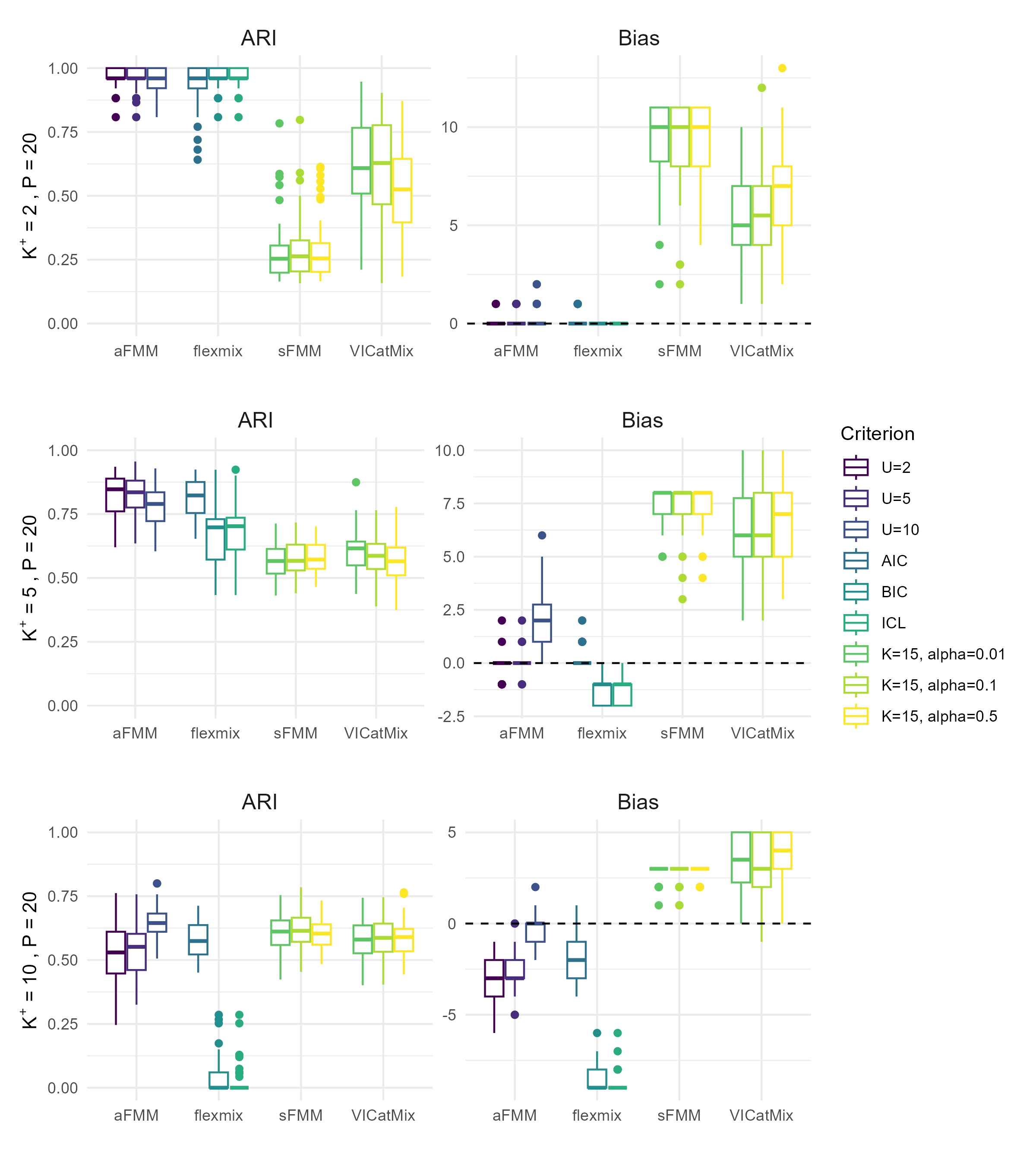}
    \caption{ARI and $K^+$ bias metrics for Scenario 1 with $K^+\in\{2,5,10\}$ and $P=20$.} \label{fig:scenario_1}
\end{figure}

\begin{figure}[H]
    \centering
    \includegraphics[width=\textwidth]{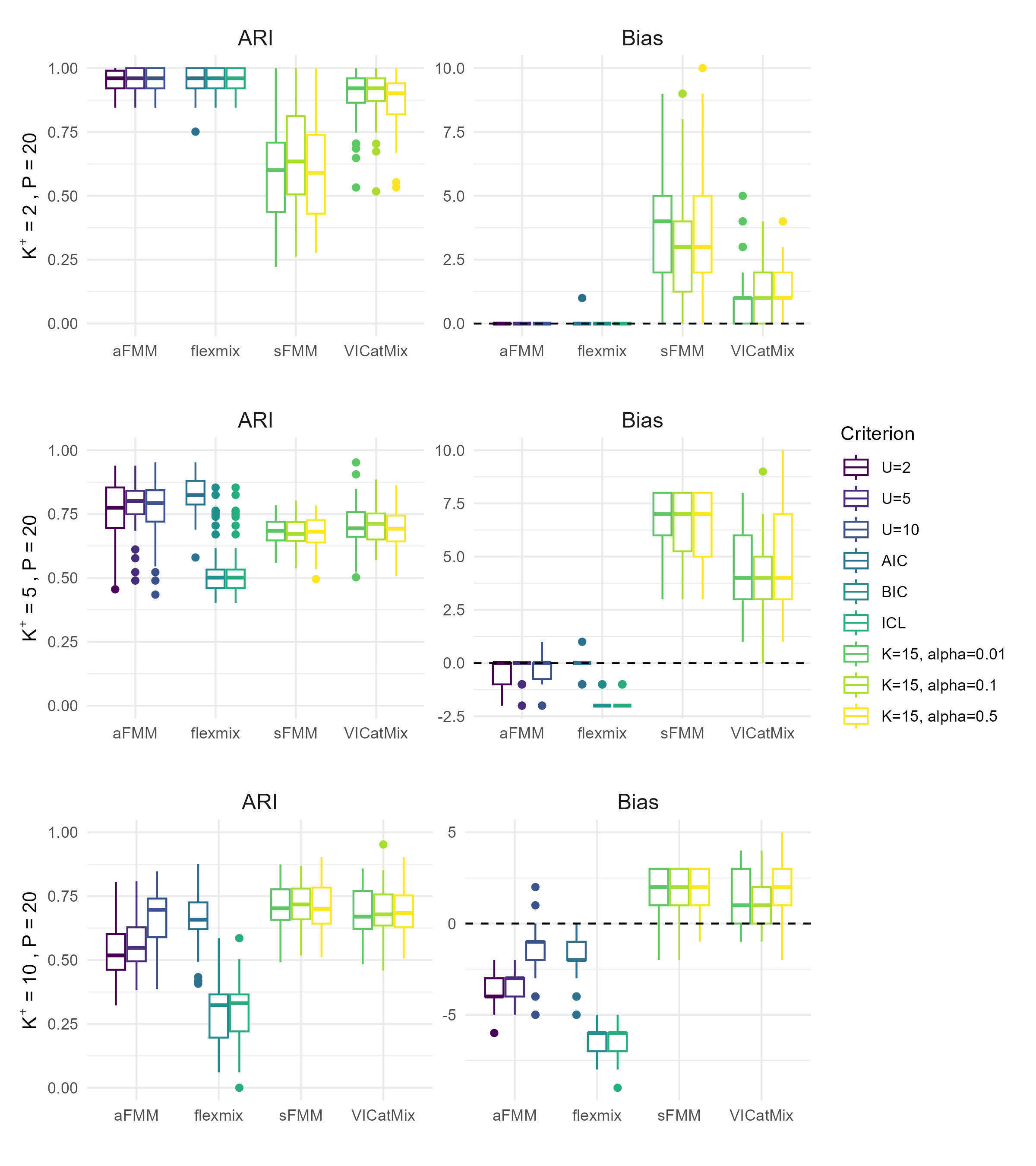}
      \caption{ARI and $K^+$ bias metrics for Scenario 2 with $K^+\in\{2,5,10\}$ and $P=20$.}
      \label{fig:scenario_2}
\end{figure}

Figures~\ref{fig:scenario_1} and \ref{fig:scenario_2} report the simulation results for both scenarios when \(P=20\). For \(K^+=2\) and \(K^+=5\), aFMM consistently outperforms both \texttt{VICatMix} and its MCMC-based counterpart sFMM in both simulation scenarios, achieving higher ARI values together with more accurate estimates of \(K^+\). This improved performance can be attributed to the use of the asymmetric Dirichlet prior, as replacing variational inference with an MCMC estimation scheme in sFMM is not sufficient to bridge the performance gap and, in fact, leads to slightly worse results. Moreover, aFMM exhibits a high degree of robustness with respect to the choice of the soft upper bound \(U\), yielding stable performance across a range of values. When \(K^+=10\), a different pattern emerges. In this setting, sFMM and \texttt{VICatMix} attain slightly higher ARI values than aFMM when \(U\) does not correspond to the number of clusters, although still systematically overestimating the number of clusters. In comparison, aFMM yields competitive clustering accuracy when \(U\) is set equal to the true number of clusters, whereas smaller values of \(U\) lead to underestimation of \(K^+\) and a corresponding reduction in ARI, as expected. 

Overall, these results indicate that aFMM provides competitive performance as long as \(U \geq K^+\). In particular, they suggest that when the number of clusters is unknown, selecting a conservatively large soft upper bound \(U\) is a sensible strategy, as it preserves clustering accuracy while maintaining robustness to prior specification. In contrast, both \texttt{VICatMix} and sFMM tend to overestimate the number of clusters consistently across different values of the shrinkage parameter \(\alpha\). This behaviour suggests that, within the symmetric Dirichlet prior framework, estimation of the number of clusters is primarily driven by the choice of the number of components \(K\), largely irrespective of the shrinkage hyperparameter \(\alpha\). Taken together, these findings suggest that aFMM offers a more reliable and robust approach.

Finally, \texttt{flexmix} yields performance comparable to that of aFMM when the estimated number of clusters is selected using the AIC criterion. However, \texttt{flexmix} only provides a partition of the observations and does not offer full posterior inference. In contrast, aFMM achieves comparable clustering accuracy while additionally providing full posterior inference, making it therefore a preferable approach in this setting.

Similar findings are obtained in the more informative case of \(P=50\) under both scenarios (see the Supplementary Material). Importantly, runtimes are broadly comparable across all four methods considered, indicating that computational cost is a negligible factor in this context. 

\section{Applications}\label{sec:applications} 
\subsection{Handwritten digits dataset}

We considered the UCI Optical Recognition of Handwritten Digits dataset and focused on the test set, which consists of $N = 1796$ labelled handwritten digit images. Each observation is represented by an $8 \times 8$ pixel grid, yielding $P = 64$ covariates. Pixel intensities in the original data are discrete; we converted the data to binary entries by assigning a value of 1 to pixels whose intensity exceeds half of the maximum possible value, and 0 otherwise. 

\begin{figure}[H]
    \centering
\includegraphics[width=\linewidth]{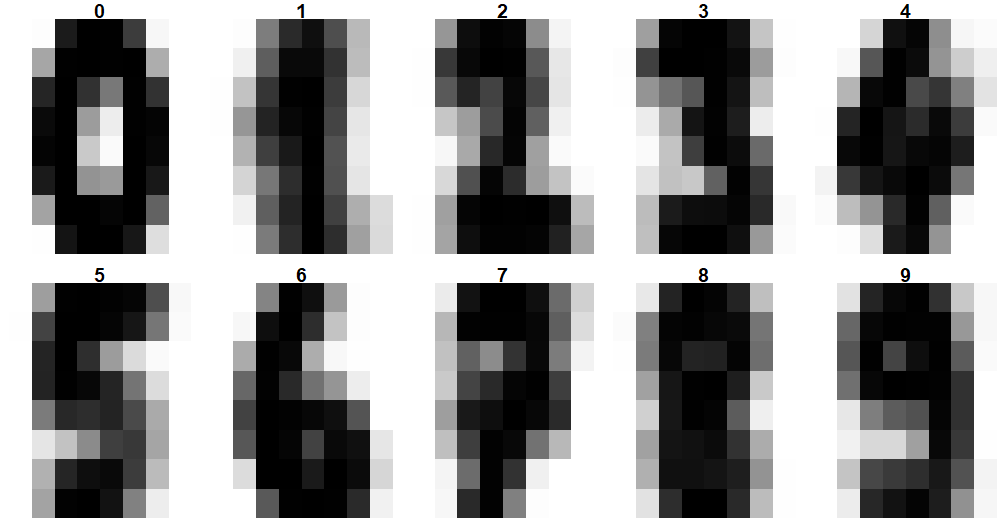}
    \caption{Mean of all the $N=1796$ binarized images grouped by digit.}\label{fig:mean_picture}
\end{figure}

Figure~\ref{fig:mean_picture} displays the averaged binarized images grouped by digit, illustrating that the distinct image patterns are preserved even after binarization.

We compared the same four approaches considered in the simulation study. First, we fitted the proposed aFMM, fixing $U = 10$ (number of digits, known \emph{a priori}) and setting $K = 15$. To assess the sensitivity of the results, we considered values of $tp \in \{0.1, 0.5, 0.9\}$. Secondly, we applied \texttt{VICatMix} with $K = 15$ and $\alpha \in \{0.01, 0.1, 0.5\}$. In addition, we considered the specification $K = 10$ and $\alpha = 1$, which fixes the number of clusters to the true value and therefore does not require shrinkage. The same models are also estimated via MCMC (sFMM). Finally, we fitted the model using \texttt{flexmix}, considering values of $K$ ranging from 5 to 15. For both \texttt{flexmix} and \texttt{VICatMix}, we used 30 initializations and exploited parallel computation on 6 cores for the latter. For the aFMM and sFMM methods, we ran 10{,}000 MCMC iterations and derived cluster assignments using the SALSO algorithm based on the variation of information criterion. 

\begin{table}
\centering
\label{tab:labels_avg}
\begin{tabular}{lcccc}
 Method  & ARI & $K^+$ estimate  & Runtime (seconds) \\ 
  \hline
flexmix (AIC, BIC, ICL) & 0.597 & 11  & 371.9 \\
\hline
VICatMix K=15 $\alpha$=0.01 &  0.515 & 15   &  545.2\\
VICatMix K=15 $\alpha$=0.1  &  0.410 & 15  & 330.3 \\
VICatMix K=15 $\alpha$=0.5  & 0.419 & 15   & 358.6\\
VICatMix K=10 $\alpha$=1    &  0.537 &  9   & 200.7 \\
\hline
sFMM K=15 $\alpha$=0.01 &  0.393 & 6 & 286.9\\
sFMM K=15 $\alpha$=0.1  &  0.636 & 15 &  229.7 \\
sFMM K=15 $\alpha$=0.5  & 0.629 & 15 & 533.5  \\
sFMM K=10 $\alpha$=1    &  0.604 &  12 &  250.9 \\
\hline
aFMM U=10 tp=0.9  & 0.633 & 15  & 312.6 \\
aFMM U=10 tp=0.5  & 0.640 & 12  & 305.2\\
aFMM U=10 tp=0.1  & \textbf{0.652} & 12  & 300.1\\
\hline
\end{tabular}
\caption{Comparison of results for the handwritten digits dataset. The $K^+$ estimation for the MCMC-based methods returns the same results using either the posterior mode or the number of clusters obtained via the SALSO algorithm.} 
\end{table}

Table \ref{tab:labels_avg} displays the results in terms of performance metrics as well as runtimes. \texttt{flexmix} provides a moderate ARI while estimating a number of clusters close to the true value, serving as a reasonable baseline. \texttt{VICatMix} generally yields lower ARI, and the estimated number of clusters is highly sensitive to the choice of $K$, even when varying the shrinkage parameter $\alpha$. The sFMM performs well as long as the shrinkage is not too strong, achieving both high ARI and cluster estimates less sensitive than \texttt{VICatMix}. The proposed aFMM is less sensitive to the choice of hyperparameter $tp$ and provides the overall best ARI when $tp = 0.1$, although it tends to slightly overestimate the number of clusters to 11. Runtimes are comparable across all methods. Overall, aFMM matches or exceeds the performance of the other methods in terms of ARI, offers runtimes comparable to competitive approaches, provides full posterior inference, and features a more intuitive framework for prior specification than sFMM.

\subsection{META$^2$ dataset}\label{sec:meta2}
\begin{figure}[H]
    \centering
    \includegraphics[width=\linewidth]{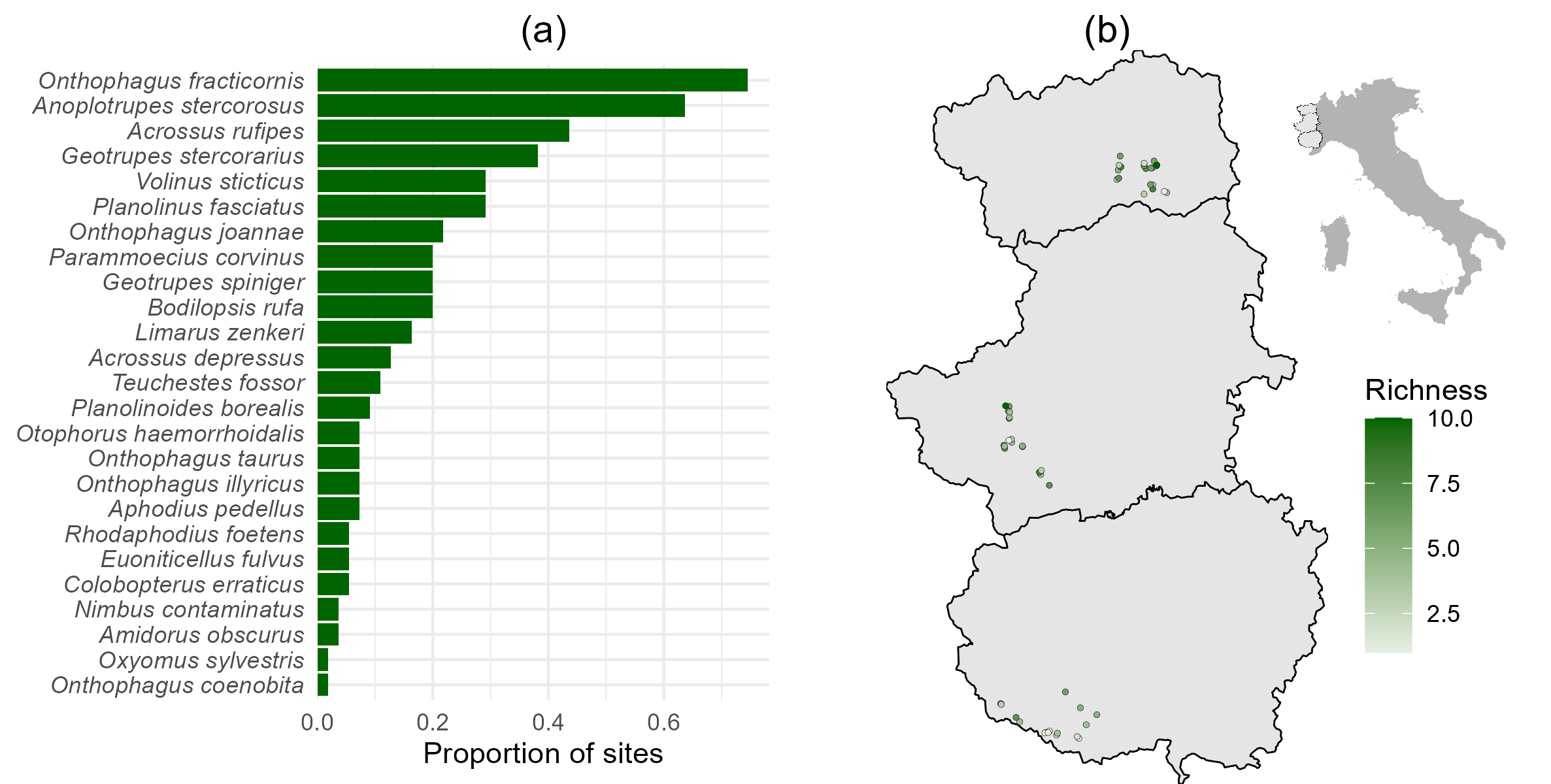}
     \caption{Overview of the META$^2$ dataset: (a) proportion of sites occupied by each species; (b) map of the Cuneo, Turin, and Aosta provinces in Italy, showing the number of species detected at each sampling location (\textit{richness}), along with a map of Italy showing the location of the provinces within the country; sites' coordinates have been slightly jittered to improve readability.}
        \label{fig:exploration}
\end{figure}

In this second application, we analyzed a metabarcoding-derived dataset from the META$^2$ project (\url{www.metasquared.unito.it}), which investigates dung-beetle biodiversity in the North-West Italian Alps ecosystem. The data report presence–absence information for $N=25$ dung-beetle species across a set of $P=55$ locations. Sampling sites are distributed across three biogeographical areas ($region\in \{\text{Graie Alps, Cozie Alps, Maritime Alps}\}$), include two land-cover types ($habitat\in\{\text{Forest, Pasture}\}$), and span three elevation bands ($elevation\in\{1000 m, 1500m, 2000m\}$). Figure~\ref{fig:exploration} provides an overview of the META$^2$ dataset.

In this application, the aim was to cluster species in order to identify groups sharing similar ecological responses to environmental gradients. To this end, environmental variables describing region, habitat, and elevation were incorporated as covariates in the species-specific success probabilities within each cluster. Formally, we considered the MBMM from Eq. \eqref{eq:MBMM} and defined:
\begin{equation}
\text{logit}(\pi_{kp}) \;=\; \beta^{(0)}_k 
\;+\; \beta^{(R)}_{k,\mathrm{region}_p}
\;+\; \beta^{(H)}_{k,\mathrm{habitat}_p}
\;+\; \beta^{(E)}_{k,\mathrm{elevation}_p},
\end{equation}
where $\beta^{(0)}_k$ are the cluster-specific intercepts capturing the baseline occurrence propensity for species in cluster $k$, while $\boldsymbol{\beta}^{(R)}_k$, $\boldsymbol{\beta}^{(H)}_k$, and $\boldsymbol{\beta}^{(E)}_k$ represent the effects associated with the environmental factors for cluster $k$. To ensure identifiability, we imposed within-cluster 0-mean constraints on the covariate coefficients, i.e. $\sum_{j=1}^3 \boldsymbol{\beta}^{(R)}_{k,j} = \sum_{j=1}^2  \boldsymbol{\beta}^{(H)}_{k,j} = \sum_{j=1}^3  \boldsymbol{\beta}^{(E)}_{k,j} = 0$ for each $k$, so that the intercepts $\beta^{(0)}_k$ correspond to the average log-odds across all factor levels.

This formulation implies that species assigned to the same cluster share a common ecological response to environmental conditions, in the sense that they are characterized by similar regression parameters. As a result, clustering will be driven by similarities in the underlying response patterns to covariates. 

\begin{figure}[H]
    \centering
    \includegraphics[width=\linewidth]{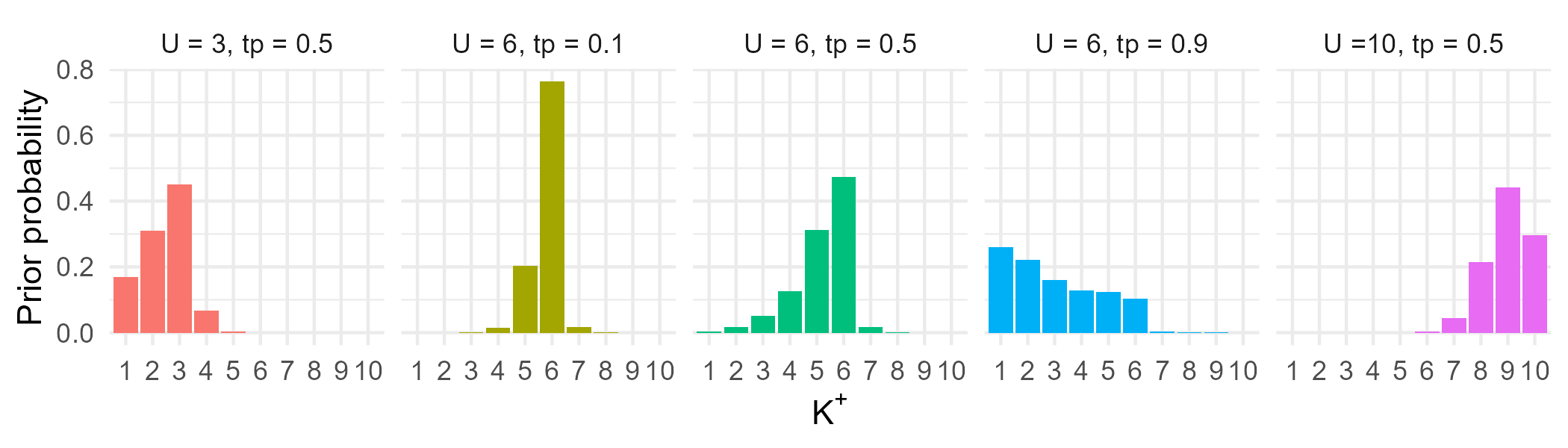}
     \caption{Induced prior distributions on the number of clusters $K^+$ for different prior choices.}
        \label{fig:sensitivity_prior}
\end{figure}

In terms of prior specification, we considered {\it iid} vague Normal priors on all $\beta$ coefficients ($\sigma^2=6.25$). As for the aFMM hyperparameters, we considered $U=6$ to be a reasonable choice based on the ecological intuition that each species may have an environmental preference among the 6 combinations of habitat $\times$ elevation levels. We considered multiple values of $tp\in\{0.1,0.5,0.9\}$ to reflect different levels of uncertainty around the number of clusters. To explore prior sensitivity, we also considered $U\in\{3,10\}$ with a moderate level of uncertainty ($tp=0.5$). The number of total components was set to $K=10$. Figure \ref{fig:sensitivity_prior} shows the variety of priors induced on $K^+$ by the different specifications. We ran all models using 50,000 iterations and retained the last 10\%.

\begin{figure}[H]
    \centering
    \includegraphics[width=\linewidth]{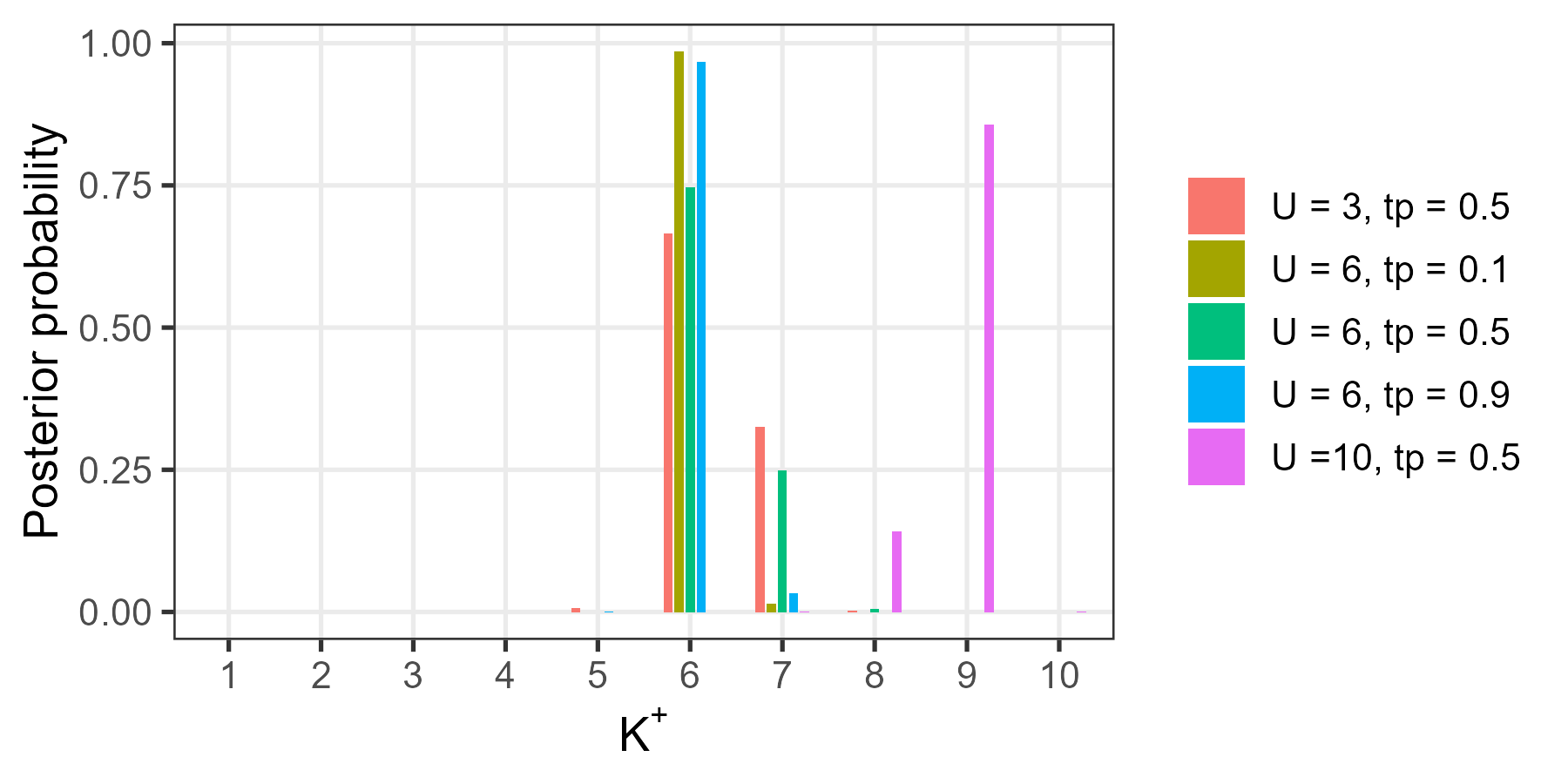}
     \caption{Posterior distribution of number of clusters  $K^+$ for different prior specifications.}
        \label{fig:sensitivity_Kplus}
\end{figure}

Figure \ref{fig:sensitivity_Kplus} shows the posterior distribution for $K^+$ for each modelling scenario. It seems that the posterior is robust to prior choice, as the posterior mode is 6 for all specifications, except the less parsimonious one ($U=10$). 

\begin{figure}[H]
    \centering
    \includegraphics[width=\linewidth]{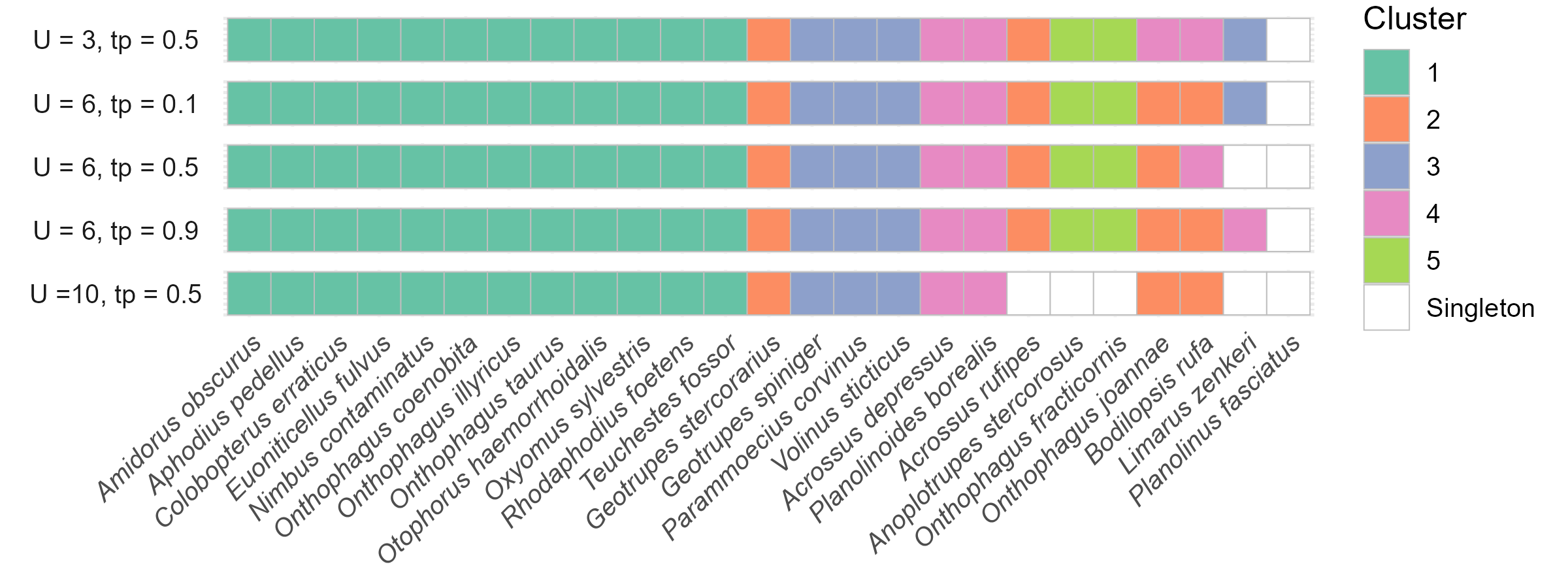} 
    \caption{Estimated partitions for different prior specifications using the SALSO algorithm based on the variation of information loss. }
        \label{fig:hard_clustering}
\end{figure}

We next examined the estimated partitions obtained from each model. To obtain these cluster assignments, we used the SALSO algorithm, again employing the variation of information loss function. Figure \ref{fig:hard_clustering} displays the cluster labels obtained under all prior specifications.

First, it can be noted that the estimated partitions are largely consistent across priors, with the exception of six species (three, if we exclude $U=10$). In general, five clusters are identified, along with between one and five singletons. From an ecological perspective, the partitions tend to separate medium-rarity level species with different environmental preferences, while maintaining the rarest and most common species in two other separate groups.

\begin{figure}[H]
    \centering
    \includegraphics[width=\linewidth]{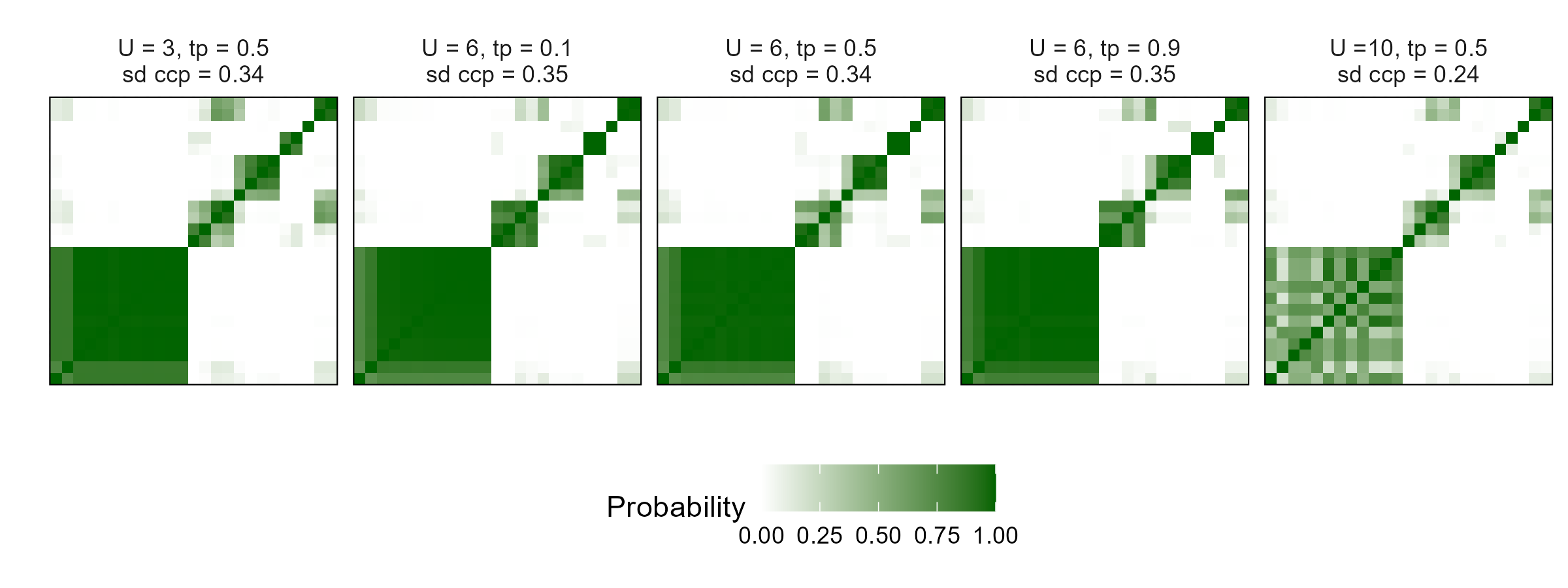}
     \caption{Posterior co-clustering matrices for different prior choices. Observations (rows and columns of the matrices) are reordered for visualization purposes in a way that is consistent across the matrices.}
    \label{fig:sensitivity_coclust}
\end{figure}

Figure \ref{fig:sensitivity_coclust} shifts the focus to soft clustering by presenting the posterior co-clustering probability matrices, together with a measure of clustering purity. A posterior co-clustering matrix is a square $N\times N$ matrix $\mathbf{C}$, whose entries are defined as $
C_{ij}
=
\Pr(Z_i = Z_j \mid \mathbf{Y})$.
As a summary measure of clustering purity, we report the standard deviation of posterior co-clustering probabilities averaged across units, computed as in \cite{page2025informed}:
\[
\mathrm{sd\;ccp}
=
\frac{1}{N}
\sum_{i=1}^N \mathrm{sd}_{j\neq i} (C_{ij}).
\]
Across prior specifications, the co-clustering matrices exhibit a well-defined block structure corresponding to six to seven clusters, in agreement with the posterior distribution of \(K^+\). The clustering structure becomes noticeably more diffuse only when \(U=10\), where the purity level drops as contrast between within-cluster and between-cluster co-clustering probabilities is reduced. Overall, these results indicate that the posterior clustering is robust to prior specification and reliably recovers the main group structures when the number of clusters is appropriately specified, whereas over-specification tends to blur the underlying structure and diminish interpretability.

Finally, we assessed posterior clustering uncertainty using the recent post-processing procedure called \textit{CHIPS} \citep{page2025}. CHIPS identifies a credible set of partitions based on a probability threshold $\gamma\in[0,1]$. It first finds the largest subpartition, i.e. a grouping of a subset of units, that appears in at least $\gamma\cdot 100\%$ of the samples. All partitions containing the subpartition then constitute the credible set. A larger size of the subpartition corresponds to a smaller credible set, indicating lower posterior uncertainty about the true partition.
For each prior specification, Table \ref{tab:chips} reports the size and posterior probability of the subpartition for $\gamma=0.5$. 
The table also reports the area under the CHIPS curve (AUChips), defined as the normalized area under the curve obtained by plotting subpartition size against subpartition probability for all values of $\gamma \in [0,1]$: AUChips provides a summary of global clustering uncertainty, ranging from 0 (complete uncertainty) to 1 (full certainty).
In all but one case, the subpartition includes at least 20 of the 25 species, with an AUChips exceeding 80\%, reflecting low uncertainty: CHIPS therefore confirms that the inferred clustering is stable, with uncertainty concentrated on a small subset of species. The exception is the extreme case of $U=10$, where both subpartition size and AUChips drop substantially, indicating greater uncertainty. The highest clustering purity is achieved with $U=6$ and $tp=0.1$. Note that the subpartitions found by CHIPS are all consistent with the partitions provided by SALSO (Figure \ref{fig:hard_clustering}).

\begin{table}[ht]
\centering
\begin{tabular}{lccc}
  \hline
 Prior & Subpartition size & Subpartition probability & AUChips \\ 
  \hline
  $U = 3, tp = 0.5$ & 20 & 0.53 & 0.83 \\ 
  $U = 6, tp = 0.1$ & 23 & 0.54 & 0.91 \\ 
  $U = 6, tp = 0.5$ & 22 & 0.57 & 0.89 \\ 
  $U = 6, tp = 0.9$ & 22 & 0.56 & 0.90 \\ 
  $U =10, tp = 0.5$ & 13 & 0.54 & 0.57 \\ 
   \hline
\end{tabular}
\caption{CHIPS metrics for different prior choices. The subpartition results are found setting the threshold $\gamma= 0.5$.}\label{tab:chips}
\end{table}

Based on ecological interpretability and metrics of clustering purity and uncertainty, we argue that the model with \(U = 6\) and \(tp = 0.1\) provides the most suitable representation of the data. 
For this model, the subpartition includes all species except \textit{Limarus zenkeri} and \textit{Nimbus contaminatus}. Unit-level uncertainty is quantified by the proportion of MCMC samples satisfying the identified subpartition in which a given unallocated unit is assigned to its modal cluster. This measure lies in $[0,1]$, with values close to 0 indicating extreme ambiguity and values close to 1 indicating full certainty. Under this criterion, \textit{Limarus zenkeri} attains a value of 0.286, while \textit{Nimbus contaminatus} reaches 0.446, indicating lower ambiguity about the cluster allocation of the latter.
The co-clustering matrix (reported in the Supplementary Material) confirms this behaviour, showing that both species exhibit substantial co-assignment probabilities across multiple clusters. In particular, they lie between the clusters to which they are assigned in Figure \ref{fig:hard_clustering} and the cluster containing \textit{Planolinoides borealis} and \textit{Acrossus depressus}.

\begin{figure}[H]
    \centering
    \includegraphics[width=\textwidth]{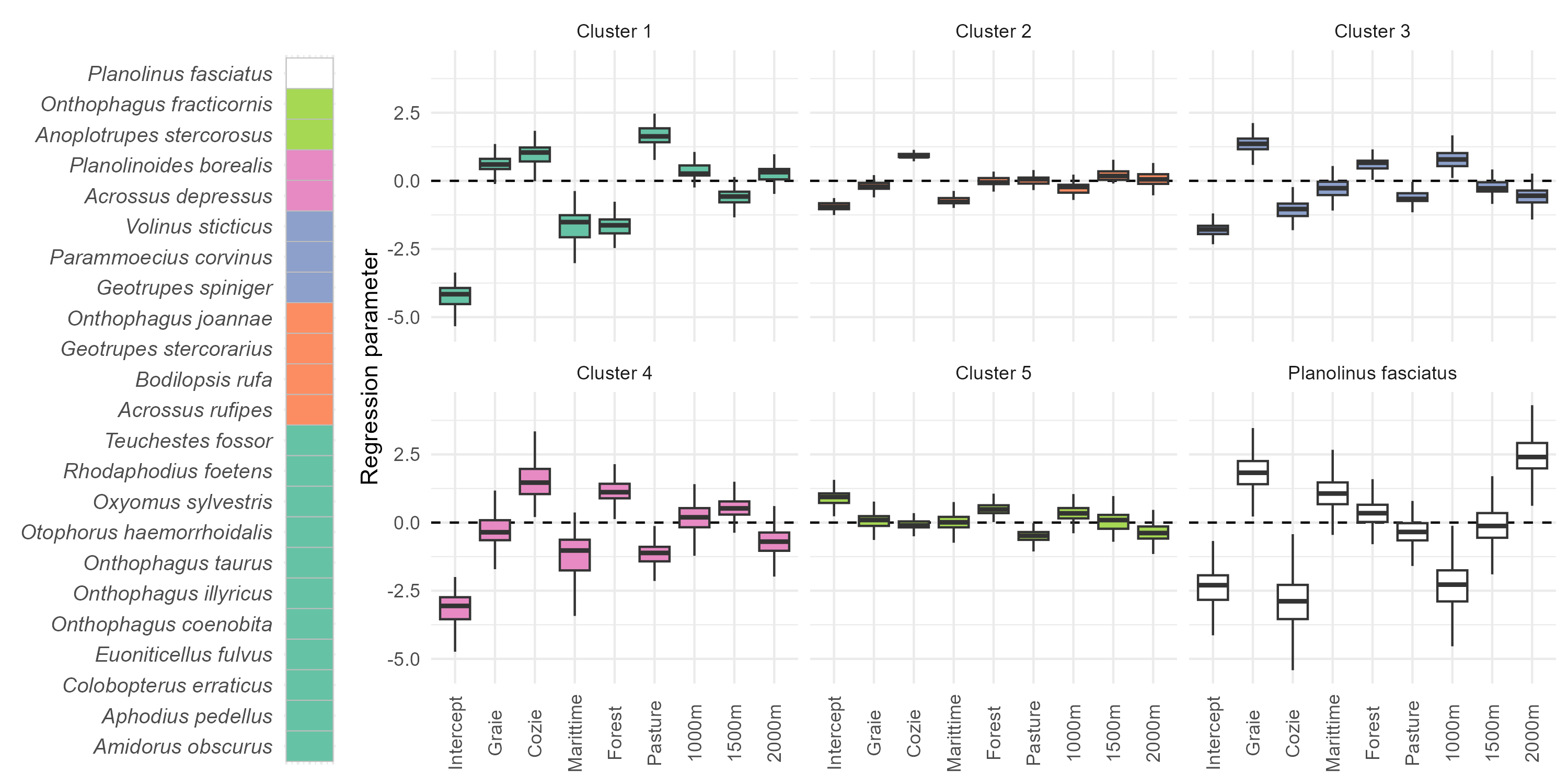} 
    \caption{Posterior distribution of the $\beta$ parameters for each cluster for the $U=6,tp=0.1$ model, conditional on the partition provided by the SALSO algorithm, represented on the left.}
        \label{fig:beta_plot}
\end{figure}

\begin{figure}[H]
    \centering
    \includegraphics[width=\textwidth]{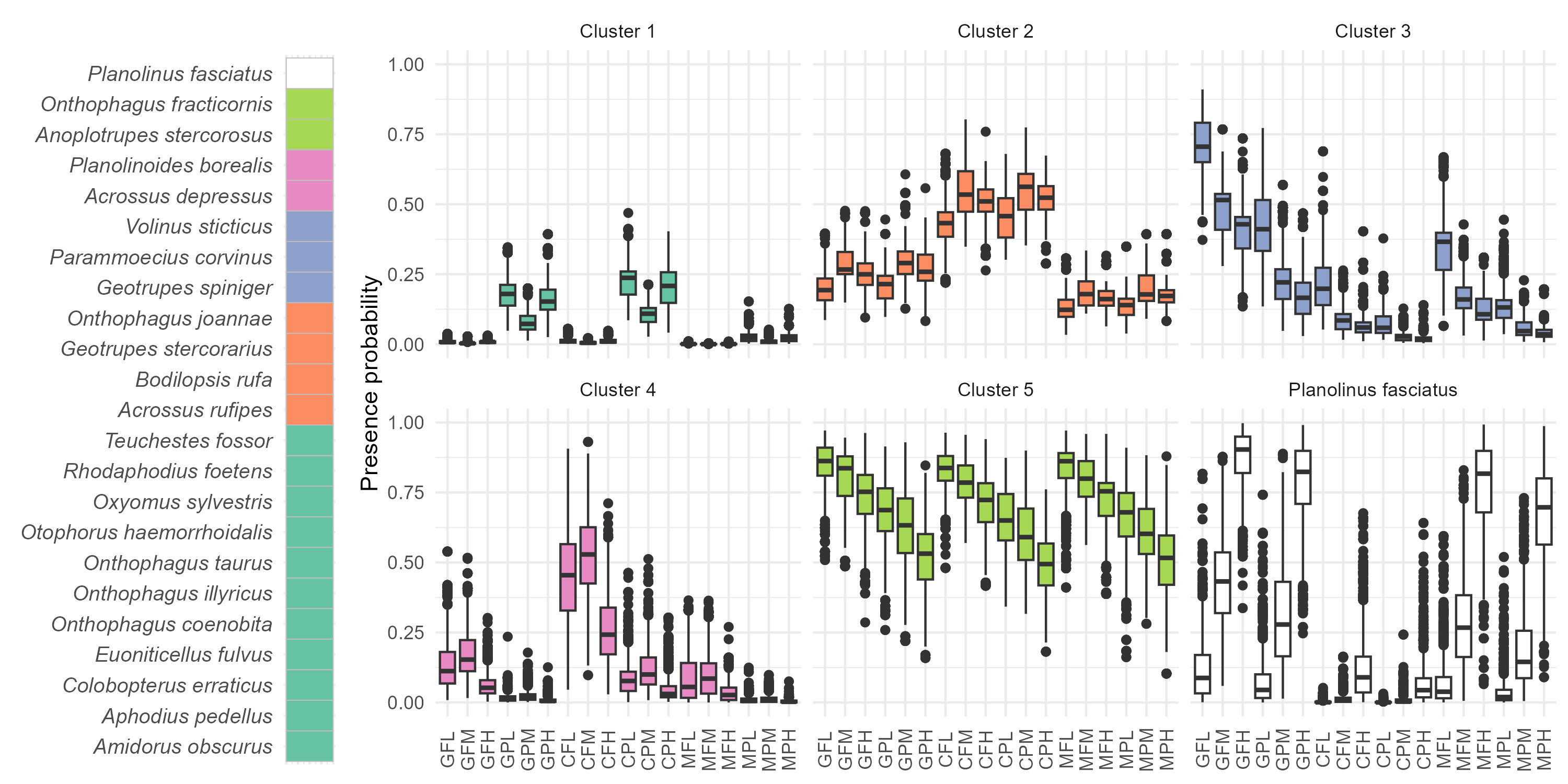} 
    \caption{Posterior distribution of the presence probabilities $\pi$ for each cluster for the $U=6,tp=0.1$ model, conditional on the CHIPS subpartition, represented on the left.  The acronyms on the x-axis correspond to the 18 possible sites defined by all combinations of covariate values: the first letter indicates the region, the second the habitat type, and the third the elevation level (Low $=1000m$, Medium $=1500m$, High $=2000m$).}
        \label{fig:probs_plot}
\end{figure}

Further insights about the selected model are provided by Figure \ref{fig:beta_plot}, which summarizes the posterior distributions of the regression parameters by cluster, and Figure \ref{fig:probs_plot}, which displays the distribution of the cluster-specific presence probabilities \(\pi_k\), evaluated at all possible covariates' combinations. The posterior distributions are obtained by considering only the posterior iterations that respect the CHIPS subpartition (2700 samples), so that they properly incorporate uncertainty about the true partition by averaging over all partitions contained in the credible set. Together, these plots provide an interpretable ecological characterization of the selected clustering solution, linking species to distinct covariate-driven occurrence patterns.

The inferred clusters allow us to draw substantive conclusions regarding species environmental preferences. Cluster 1 comprises rare species predominantly associated with pasture habitats in the Graie and Cozie Alps, with no clear pattern along the elevation gradient. Cluster 2 groups relatively common species that are most prevalent in the Cozie region, followed by the Graie Alps, and exhibit no marked preference for habitat type. Species in Cluster 3 show a clear preference to low to mid elevations and forested environments, particularly within the Cozie Alps, whereas Cluster 4 displays a similar preference for forest habitats at low elevations but with greater prevalence in the Graie region. Cluster 5 includes the two most common species in the dataset, characterized by a broad distribution across all Alpine regions and habitat types, with a tendency toward lower elevations and forest environments. Finally, \textit{Planolinus fasciatus} is singled out mainly by its preference for high-elevation sites, a slight preference for forest habitats, and a notably lower occurrence in the Cozie Alps. 
Overall, the identified environmental associations are broadly consistent with patterns reported in previous ecological studies on alpine dung beetles \citep{macagno2009maintenance,negro,environments11080178}. A notable deviation concerns \textit{Onthophagus fracticornis}, which, although typically associated with pasture habitats, shows a weak preference for forest environments in our analysis.
 
\section{Discussion}\label{sec:discussion} 
 In this paper, we proposed the use of the recently introduced asymmetric Dirichlet prior within the framework of MBMMs as a competitive alternative for model-based clustering of multivariate binary data. Our approach offers several key advantages that can be achieved simultaneously, whereas existing methods typically provide only some of these benefits at a time. Specifically, it delivers full posterior inference (being a Bayesian implementation), attains high computational efficiency through a sparse finite mixture specification that does not require trans-dimensional MCMC, and provides a transparent, user-friendly way to encode prior knowledge about the number of clusters via the Penalized Complexity framework, facilitating the interpretation of prior sensitivity.  Simulation studies confirmed these concurrent benefits across scenarios with varying levels of informativeness. Similar patterns were observed in the application to the handwritten digits dataset.

 The second application illustrated an extension of the framework to the incorporation of covariates. To our knowledge, this is the first MBMM application allowing covariates indexed by the $P$ dimension to inform clustering. This extension is particularly relevant in the context of DNA metabarcoding \citep{taberlet2018environmental}, which increasingly produces larger presence-absence datasets, compared to those datasets obtained via morphological identification. The higher dimensionality of these new datasets makes a joint modelling approach, such as the one of the \texttt{Hmsc} package by \cite{tikhonov2020joint}, computationally intensive: this MBMM extension offers a practical alternative, modelling the data via mixtures rather than by estimating the full correlation structure between units, while still allowing for the inclusion of environmental covariates. Future work will explore this approach on larger metabarcoding datasets to assess both predictive performance and ecological relevance. While our applications focused on ecological and image data, the proposed method is broadly applicable to any domain involving multivariate binary data, including biomedicine, social policy, text classification, voting behavior, and animal trait studies.

Several limitations of the proposed approach should be acknowledged. First, the assumption of independence, or conditional independence given covariates, may not hold in some applications, suggesting the need for more complex regression structures, such as spatial or temporal effects, when the $p$-dimension suggests it. Second, careful tuning of the MCMC algorithm is required, particularly for the update of the $\alpha_1$ parameter. Also, the computational burden may limit scalability when $P$ is large or when the number of fixed and random effects increases. To address this, future work could explore improvements to the MCMC algorithm, potentially by combining it with alternative inferential approaches such as integrated nested Laplace approximation \citep{gomez2018markov}. Third, the method could be extended beyond binary outcomes, to the clustering of multivariate categorical data. Finally, this work, together with the results reported in \cite{page2025informed}, confirms that the asymmetric Dirichlet prior performs well in both Gaussian and multivariate Bernoulli mixture models, suggesting that its use in other GLM-based mixtures is a promising direction for future research.

\section*{Declarations}
\textbf{Funding:} LF, MFV, and AL were funded by the European Union under the NextGeneration EU Programme within the Plan “PNRR - Missione 4 “Istruzione e Ricerca” - Componente C2 Investimento 1.1 “Fondo per il Programma Nazionale di Ricerca e Progetti di Rilevante Interesse Nazionale (PRIN)” by the Italian Ministry of University and Research (MUR), Project title: “METAbarcoding for METAcommunities: towards a genetic approach to community ecology (META2) ”, Project code: 2022PA3BS2 (CUP E53D23007580006), MUR D.D. financing decree n. 1015 of 07/07/2023. \\ 
LF and MFV were also partly funded by the University of Modena and Reggio Emilia Departmental Research Funding (FAR dipartimentale 2025) PROT. 140157 of 27/05/2025.

\textbf{Competing Interests:} The authors have no competing interests to declare that are relevant to the content of this article.  

\textbf{Availability of Data and Materials:} Data and materials are available from the corresponding author upon reasonable request.  

\textbf{Code Availability:} Code to replicate the simulation study and the two applications is publicly available at \url{https://github.com/LFerrariIt/binary_clustering}.

\textbf{Authors' Contributions:} LF, MFV, and GP conceptualized the article. LF carried out the coding tasks. AL provided the ecological dataset and expert guidance in its interpretation and analysis. All authors contributed to methodology, discussion of results, and manuscript writing, and approved the final version of the manuscript.

\paragraph{Acknowledgements} We thank Nicolò Brunelli, Daniele Bertolino, Andrea Pelle, Laura Gruppuso, Samuele Voyron, Angela Roggero and Claudia Palestrini for their contribution in the collection and processing of the dataset used in the ecological application.
\newpage

\bibliographystyle{plainnat}

\bibliography{sn-bibliography}

\newpage

\begin{center}
{\Large
\textbf{Supplementary Material for ``Informed Asymmetric Dirichlet Priors for Multivariate Bernoulli Mixture Models''}} \\
\vspace{0.3cm}
\textit{Luisa Ferrari, Maria Franco Villoria, Garritt L. Page, Alex Laini}
\end{center}

\setcounter{figure}{0}
\renewcommand{\thefigure}{S\arabic{figure}}

\section*{S1. Visualization of the simulated datasets}

The following figures illustrate the datasets considered in Scenarios 1 and 2. For visualization purposes, observations have been ordered according to their cluster labels.
\begin{figure}[H]
\centering
\includegraphics[width=0.7\textwidth]{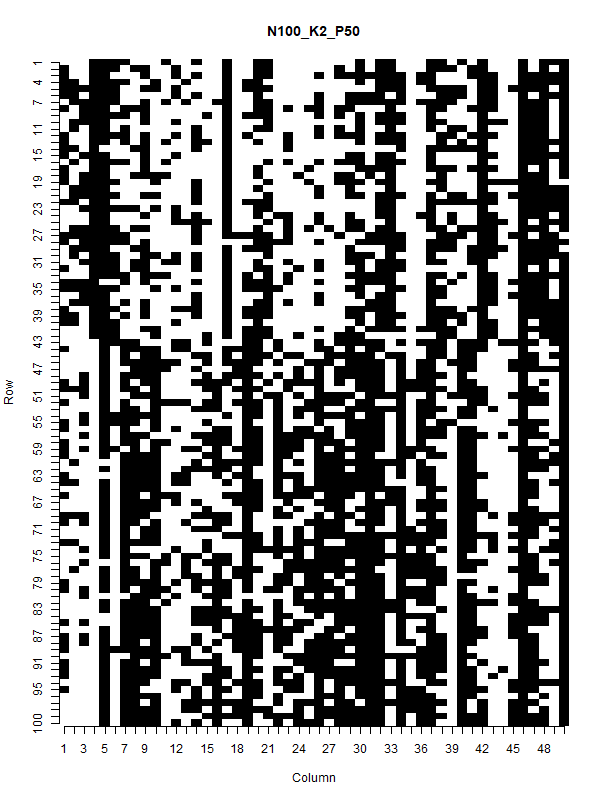}
\caption{Scenario 1 dataset with $K^+=2$ and $P=50$. Each tile represents a binary observation; rows correspond to observations ordered by true cluster membership.}
\end{figure}

\begin{figure}[H]
\centering
\includegraphics[width=0.7\textwidth]{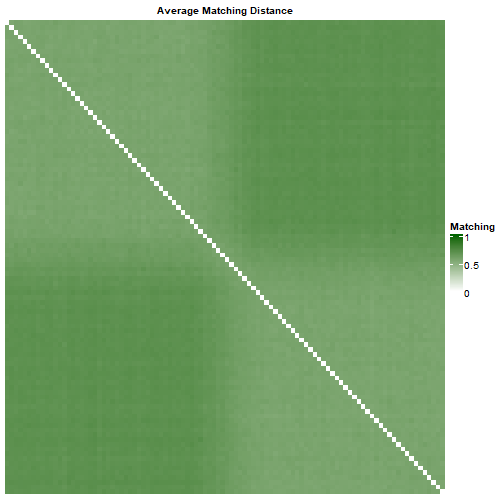}
\caption{Scenario 1 dataset with $K^+=2$ and $P=50$: matching distance matrix between observations averaged across the 50 datasets.}
\end{figure}

\begin{figure}[H]
\centering
\includegraphics[width=0.7\textwidth]{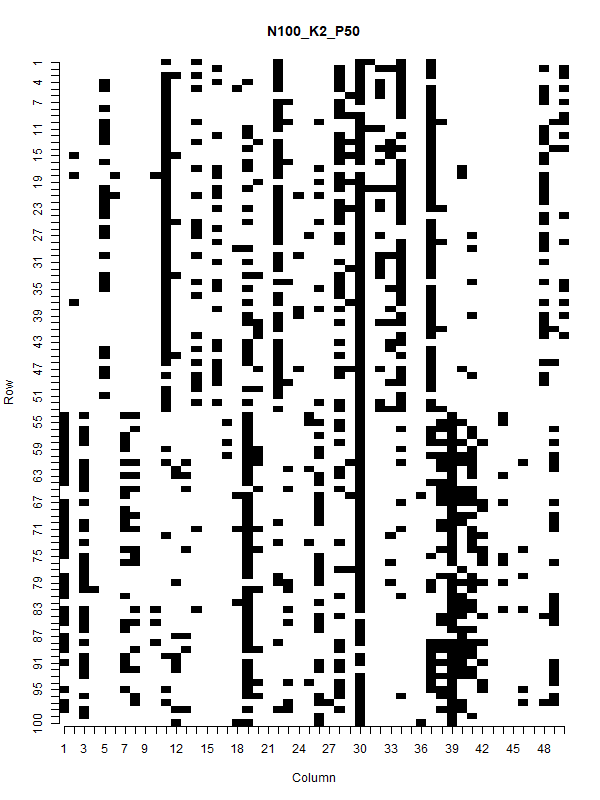}
\caption{Scenario 2 dataset with $K^+=2$ and $P=50$. Each tile represents a binary observation; rows correspond to observations ordered by true cluster membership.}
\end{figure}

\begin{figure}[H]
\centering
\includegraphics[width=0.7\textwidth]{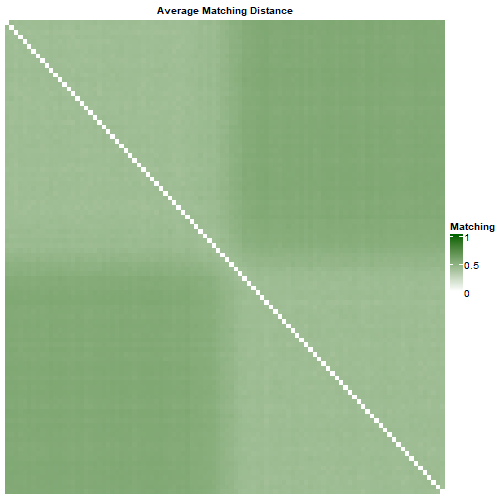}
\caption{Scenario 2 dataset with $K^+=2$ and $P=50$: matching distance matrix between observations averaged across the 50 datasets.}
\end{figure}

\section*{S2. Additional simulation results}

The following figures illustrate more results from the simulations study, specifically the results of Scenario 1 and 2 in the case in which $P$ is set to 50.
\begin{figure}[H]
    \centering
    \includegraphics[width=\textwidth]{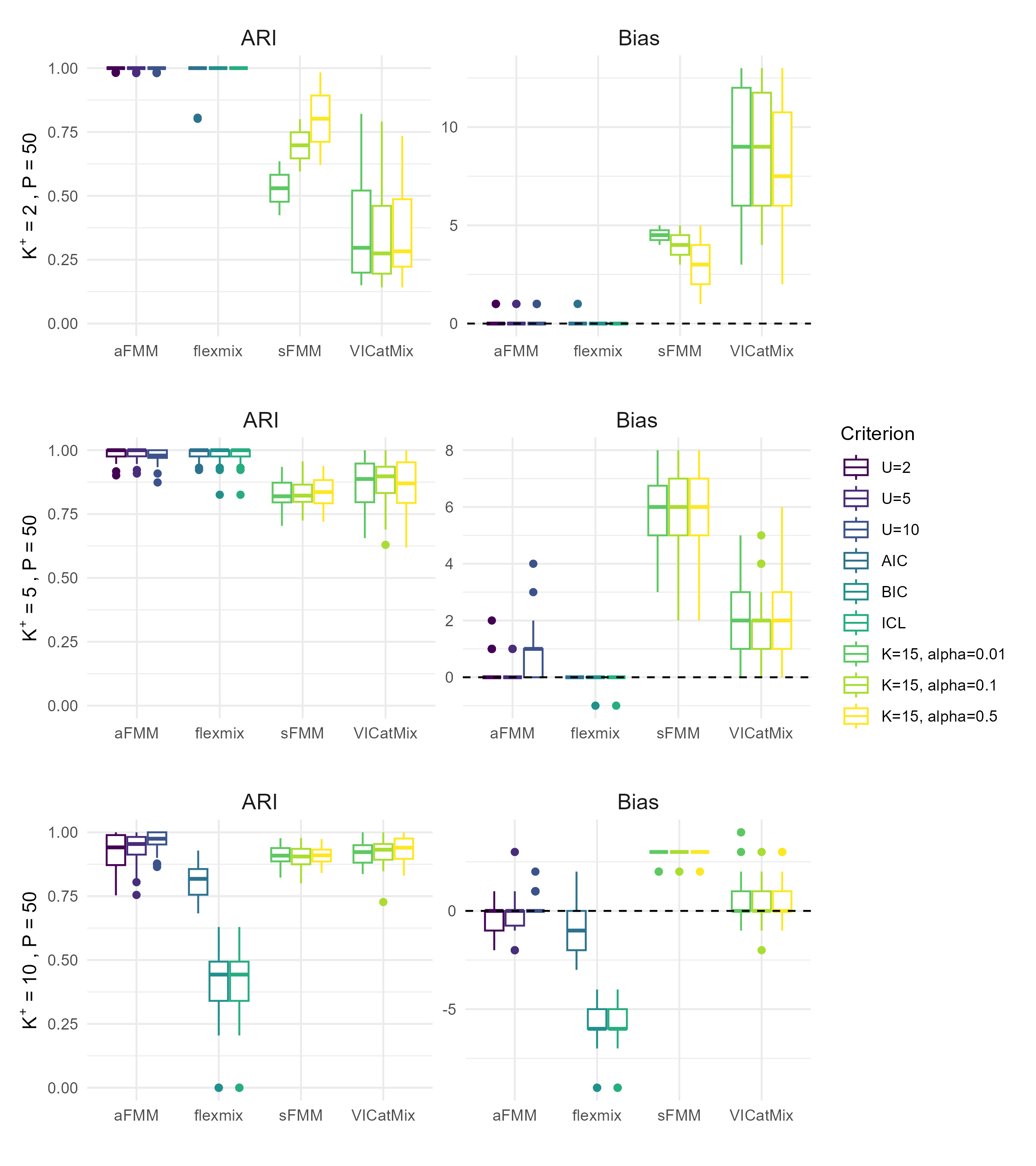}
    \caption{ARI and $K^+$ bias metrics for Scenario 1 with $K^+\in\{2,5,10\}$ and $P=50$.} \label{fig:scenario_1_sm}
\end{figure}

\begin{figure}[H]
    \centering
    \includegraphics[width=\textwidth]{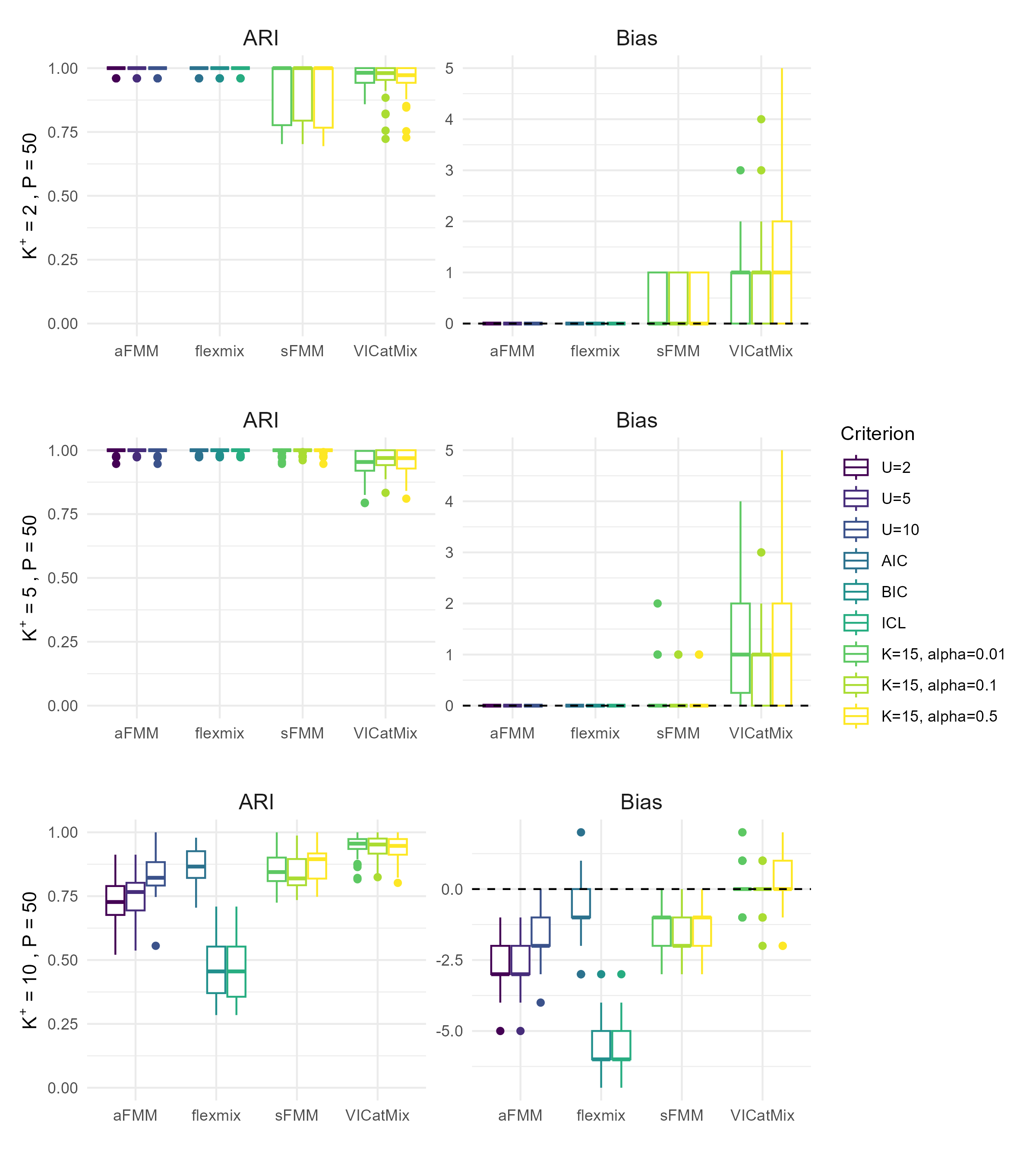}
      \caption{ARI and $K^+$ bias metrics for Scenario 2 with $K^+\in\{2,5,10\}$ and $P=50$.}
      \label{fig:scenario_2_sm}
\end{figure}


\section*{S3. Additional results on the META$^2$ dataset application}

\begin{figure}[H]
    \centering
    \includegraphics[width=\textwidth]{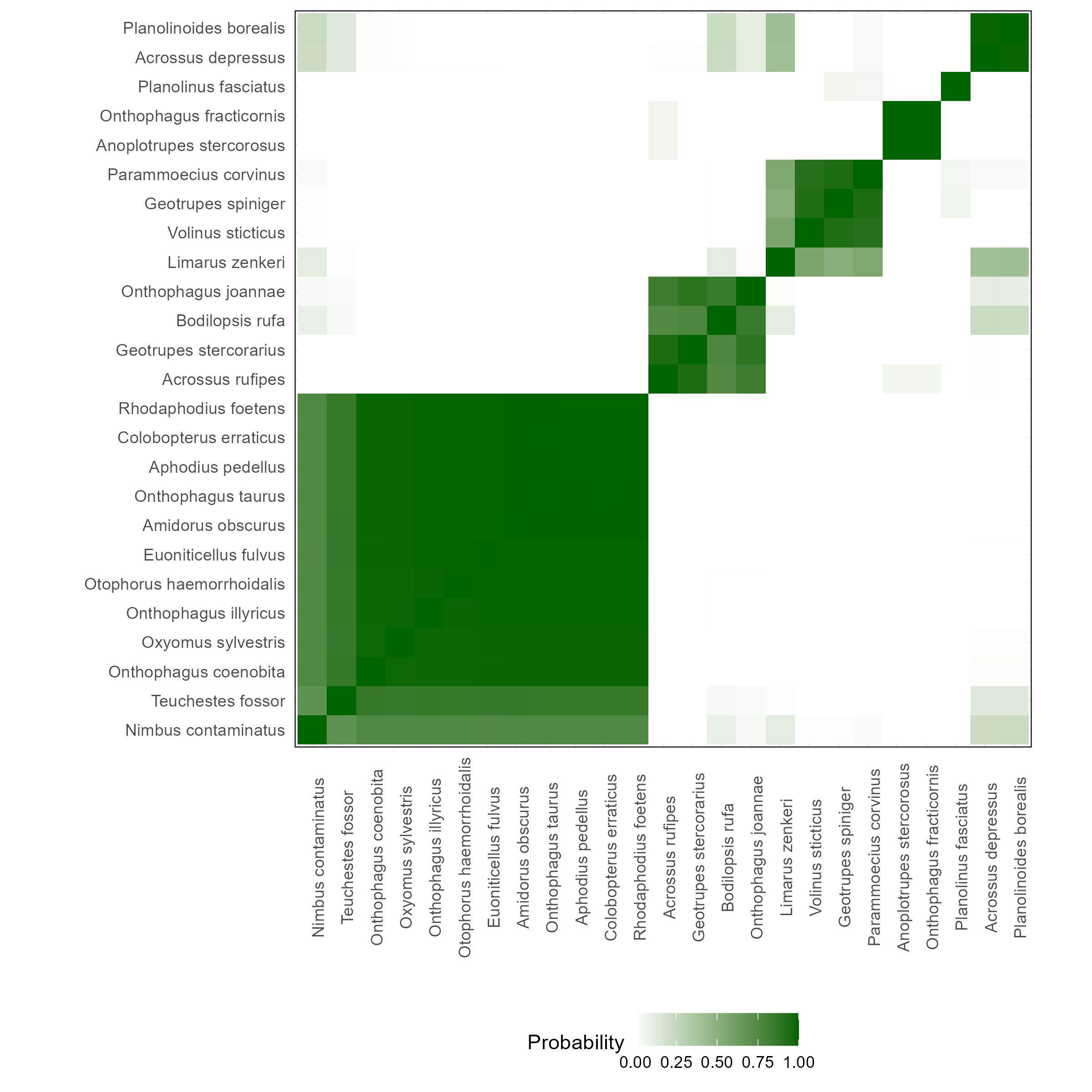}
      \caption{Posterior co-clustering matrix for the model $U=6,tp=0.1$.}
      \label{fig:psm_sm}
\end{figure}

\section*{S4. MCMC algorithm code}

The following R code implements the MCMC algorithm described in the main text; additional functions and supporting scripts are available in the associated GitHub repository: \url{https://github.com/LFerrariIt/binary_clustering/}.

\vspace{0.5cm}

\begin{lstlisting}
aFMM_function <- function(
    Y,                      # N x P binary response matrix
    U, tp, K, alpha2=0.01,  # aFMM hyperparameters
    alpha1_density,         # PC prior density function for alpha1 (precomputed)
    a=0.5, b=0.5,           # Beta(a,b) prior parameters for Bernoulli probabilities
    proposal_sd_alpha1 = 1, # sd for RW Metropolis for alpha1
    n_iter = 10000          # total number of MCMC iterations
) {
  
  ##############################
  # Basic dimensions
  ##############################
  N <- nrow(Y)   # number of observations
  P <- ncol(Y)   # number of binary variables
  
  ##############################
  # Temperature schedule for simulated annealing
  ##############################
  temperature <- c(
    # cooling phase
    exp(seq(log(5), log(1), length.out = round(n_iter * 9/10))),  
    # standard MCMC
    rep(1, round(n_iter / 10))                                    
  )
  
  ##############################
  # Log-posterior for alpha1
  ##############################
  # Computes log p(alpha1 | omega, U) up to normalizing constant
  log_post_alpha1 <- function(a1, omega, U) {
    log_lik <- lgamma(U * a1) - U * lgamma(a1) +
      (a1 - 1) * sum(log(omega[1:U]))
    prior <- log(alpha1_density(a1))
    return(log_lik + prior)
  }
  
  ##############################
  # Storage objects
  ##############################
  z_samples <- matrix(NA, n_iter, N)
  pi_samples <- array(NA, dim = c(n_iter, K, P))
  
  ##############################
  # Initialization
  ##############################
  # initial cluster labels using k-modes 
  z <- klaR::kmodes(lapply(as.data.frame(Y), as.factor), modes = U)$cluster
  alpha1 <- 1
  omega <- MCMCpack::rdirichlet(1, c(rep(alpha1, U), rep(alpha2, K - U)))
  pi <- matrix(rbeta(K * P, a, b), nrow = K, ncol = P)
  
  ##############################
  # MCMC main loop
  ##############################
  for (iter in 1:n_iter) {
    
    ##############################
    # STEP 1: Update cluster allocations z_i
    ##############################
    # Compute log-probability of each observation belonging to each cluster
    log_pi <- log(pi)               # log Bernoulli success probabilities
    log1m_pi <- log(1 - pi)         # log Bernoulli failure probabilities
    log_omega <- log(omega)         # log mixture weights
    # log-likelihood of each observation under each cluster
    log_lik <- matrix(log_omega, nrow(Y), length(omega), byrow = TRUE) +
      Y %*% t(log_pi) +
      (1 - Y) %*% t(log1m_pi)
    # apply annealing temperature
    log_lik_temp <- log_lik / temperature[iter]
    # numerical stabilisation (subtract row max)
    log_lik_temp <- log_lik_temp - matrixStats::rowMaxs(log_lik_temp)
    # convert to probabilities
    prob_mat <- exp(log_lik_temp) / rowSums(exp(log_lik_temp))
    # sample new cluster labels 
    z <- apply(prob_mat, 1, function(p) sample(1:K, 1, prob = p))
    # re-order the cluster labels based on decreasing cluster size
    n_k <- tabulate(z, nbins = K)
    cluster_order <- order(n_k, decreasing = TRUE)
    z <- match(z, cluster_order)  
    
    ##############################
    # STEP 2: Update mixture weights omega
    ##############################
    # Dirichlet update using cluster counts
    alpha_vec <- c(rep(alpha1, U), rep(alpha2, K - U))
    omega <- MCMCpack::rdirichlet(1, alpha_vec + tabulate(z, nbins = K))
    
    ##############################
    # STEP 3: Update Bernoulli parameters pi
    ##############################
    for (k in 1:K) {
      indices <- which(z == k)  # observations in cluster k
      n_k <- length(indices)    # cluster size
      y_sum <- colSums(Y[indices, , drop = FALSE])  # successes per variable
      pi[k, ] <- rbeta(P,a + y_sum,b + n_k - y_sum)
    }
    
    ##############################
    # STEP 4: Update alpha1 (Metropolis step)
    ##############################
    # random walk proposal
    prop_alpha1 <- rnorm(1, alpha1, proposal_sd_alpha1)
    # enforce support constraints for alpha1
    if (prop_alpha1 > 0.05 & prop_alpha1 <= U) {
        log_post_curr <- log_post_alpha1(alpha1, omega, U)
        log_post_prop <- log_post_alpha1(prop_alpha1, omega, U)
        # Metropolis acceptance step
        if (log(runif(1)) < (log_post_prop - log_post_curr)) {
          alpha1 <- prop_alpha1
        }
      }
    
    ##############################
    # STEP 5: Store samples
    ##############################
    z_samples[iter, ] <- z
    pi_samples[iter, , ] <- pi
  }
  
  ##############################
  # Posterior summaries 
  ##############################
  # Remove burn-in period
  burnin <- round(n_iter * 9/10)
  trimmed_z_samples <- z_samples[burnin:n_iter, ]
  trimmed_pi_samples <- pi_samples[burnin:n_iter, , ]
  
  ##############################
  # Output
  ##############################
  return(list(
    "z_samples" = trimmed_z_samples,
    "pi_samples" = trimmed_pi_samples
  ))
}
\end{lstlisting}

\end{document}